\documentclass[aps,prd,twocolumn,superscriptaddress,nofootinbib,floatfix,longbibliography]{revtex4-2}

\usepackage[T1]{fontenc}
\usepackage[utf8]{inputenc}
\usepackage{amsmath, amssymb, amsfonts}
\usepackage{graphicx}
\usepackage{xcolor}
\usepackage[colorlinks=true, citecolor=blue, linkcolor=blue, urlcolor=blue]{hyperref}
\usepackage{bm}
\numberwithin{equation}{section}

\newcommand{\dd}{\mathrm{d}}

\newcommand{\mchi}{m_\chi}

\newcommand{\p}{'}

\newcommand{\degree}{^\circ}

\usepackage{tcolorbox}
\usepackage{mdframed}

\begin{document}

\title{Photon--Dark Matter Elastic Scattering: An Effective-Operator Scan and First Operator-Resolved Sensitivity Estimates from the Galactic Halo}
\author{T.~R.~Stenhouse}
\email{trinity.stenhouse.24@ucl.ac.uk}
\affiliation{Department of Physics and Astronomy, University College London, Gower Street, London WC1E 6BT, United Kingdom}
\author{A.~Acar}
\affiliation{School of Physics, Engineering and Technology, University of York, Heslington, York YO10 5DD, United Kingdom}
\author{M.~Bashkanov}
\affiliation{School of Physics, Engineering and Technology, University of York, Heslington, York YO10 5DD, United Kingdom}
\author{F.~F.~Deppisch}
\affiliation{Department of Physics and Astronomy, University College London, Gower Street, London WC1E 6BT, United Kingdom}
\author{C.~Ghag}
\affiliation{Department of Physics and Astronomy, University College London, Gower Street, London WC1E 6BT, United Kingdom}
\author{D.~P.~Watts}
\affiliation{School of Physics, Engineering and Technology, University of York, Heslington, York YO10 5DD, United Kingdom}
\date{\today}

\begin{abstract}
Elastic photon--dark matter scattering attenuates gamma-ray
spectra along a line of sight, probing the same operators as dark matter annihilation to
photons but at a rate linear, rather than quadratic, in dark matter density. We consider Standard Model gauge-invariant effective operators of mass-dimension 5 to 7,
suppressed by a cutoff scale $\Lambda$, coupling scalar, Majorana or Dirac dark matter
to the photon. The leading operators with non-vanishing real-photon amplitudes enter at
dimension-5 for Dirac dark matter and dimension-7 for Majorana dark matter. In the electroweak-doublet dipole portal, the inelastic splitting
invoked to evade direct detection also closes the CMB annihilation bound, leaving attenuation the only one of the three photon-sector probes that survives. Applying this to a pixel-level reanalysis of 17 years of
\textit{Fermi}--LAT Pass~8 data toward the Galactic centre, we derive the first operator-resolved sensitivity estimates for photon--dark matter scattering from the Galactic halo: $\Lambda \simeq 0.32~\mathrm{GeV}$ for the dimension-5 Dirac dipoles,
$0.21~\mathrm{GeV}$ for the dimension-6 scalar Rayleigh operator and
$0.79$--$1.06~\mathrm{GeV}$ for the dimension-7 Rayleigh family. The reach is weak: it lies below the EFT-validity threshold across the cold dark matter mass range, and is superseded on the dipole plane by CMB and direct-detection constraints. The
framework is calibrated against pseudo-experiments, and recomputes the sensitivity for any instrument that provides a per-bin spectrum with uncertainties and a
line-of-sight column density.
\end{abstract}

\maketitle

% Your body text will go here.
%==================================================================
\section{Introduction}
\label{sec:introduction}
%==================================================================

Dark matter (DM) accounts for roughly 85\% of the matter content of the universe. Its gravitational imprint is unambiguous in galaxy rotation curves~\cite{rubin1970rotation}, gravitational lensing~\cite{Massey_2010}, and the temperature anisotropies of the cosmic microwave background~\cite{Planck2018CosmoParams}, yet no direct measurement has isolated the particle carrying that mass. The conventional experimental programme rests on three complementary channels: direct detection of DM--nucleon recoils in underground laboratories~\cite{XENONnT2025}, collider searches for missing-energy signatures~\cite{Kahlhoefer2017CollderReview,Boveia2018CollDMWWG}, and indirect detection of Standard Model (SM) final states produced by DM annihilation or decay in astrophysical environments~\cite{Slatyer2021IndirectReview,Gaskins2016IndirectReview}. Each channel presupposes a particular class of DM--SM interaction, and each has so far returned null results across broad swathes of parameter space.

A complementary handle on the same interactions, largely underexplored, is the elastic scattering of SM photons off DM in the Galactic halo or along cosmological lines of sight. 
No tree-level photon--DM vertex exists in the SM for any neutral DM candidate, but loop-induced couplings through charged intermediaries generate a gauge-invariant amplitude. 
The operators involved are not new: the same vertices generate DM annihilation to photons, $\chi\bar\chi\to\gamma\gamma$, already constrained by gamma-ray line searches~\cite{KavanaghPanciZiegler2019FaintLight,Barducci2025ScalarRayleigh}, and pair production $\gamma\gamma\to\chi\bar\chi$, which colliders probe through vector-boson-fusion and mono-$X$ final states~\cite{Arina2021PhotonEFT}. 
Elastic scattering is the crossed channel of those processes, probing the same operator basis in a different kinematic regime: at spacelike momentum transfer with both photons on shell, rather than at $s = 4m_\chi^{2}$, and with a rate linear rather than quadratic in DM density. 
The difference is not merely technical; it changes which operators contribute at all. The anapole and charge-radius structures have identically vanishing real-photon elastic amplitudes (Sec.~\ref{subsec:eft_catalogue}), while remaining constrained by direct detection and colliders.
Below the scale of the charged intermediaries, this coupling is captured by a set of higher-dimension effective operators built from the DM bilinear and the photon field strength (Sec.~\ref{subsec:eft_catalogue}). 
Each carries a different energy dependence of the elastic cross section: for
$\omega \ll m_\chi$, $\sigma \propto E_\gamma^{2}$ for the dimension-5 dipoles
and $E_\gamma^{4}$ for the dimension-6 scalar Rayleigh operator, while the
dimension-7 fermionic Rayleigh operators split, $E_\gamma^{4}$ for the
parity-even structure and $E_\gamma^{6}$ for the parity-odd one.
The shape a spectrum acquires through this attenuation and energy redistribution therefore depends on the scattering operator, not merely its normalisation, and leaves a characteristic signature on the diffuse and point sources observed by current gamma-ray instruments, notably the \textit{Fermi} Large Area Telescope (\textit{Fermi}--LAT)~\cite{Atwood2009FermiLAT}.

At the CMB epoch, photon--DM scattering is bounded by Planck data~\cite{Wilkinson2014DMphotonCMB,Stadler2018DMphotonPlanck}. In the late universe, the only line-of-sight constraint is that of Ref.~\cite{acar2026darkmatterredblue}, which treats two specific ultraviolet (UV) completions, a SM Higgs portal and perturbative quantum gravity, with a heavy weakly interacting massive particle (WIMP) as the DM candidate. Ref.~\cite{acar2026darkmatterredblue} does not set a limit on $\mchi$, but argues that non-observation of $\gamma\chi\to\gamma\chi$ could in principle bound heavy gravitationally interacting candidates, and notes that any such bound is likely weaker than existing microlensing constraints. The bound is tied to the two UV completions rather than to an operator basis, so does not translate directly onto an operator-independent DM-mass exclusion, and detectability is set by a fixed optical-depth threshold rather than by the statistical uncertainty of the underlying spectrum, which would yield a stronger and continuous figure of merit.

The paper is organised as follows. Sec.~\ref{sec:methodology} sets out the
attenuation and reshaping formalism, the test statistic and its calibration, and
the datasets. Sec.~\ref{sec:UV_revisit} revisits the two UV completions of
Ref.~\cite{acar2026darkmatterredblue} and finds both unobservable.
Sec.~\ref{sec:eft_operators} develops the dimension-5 through -7 operator scan
and its validity criteria, Sec.~\ref{sec:halo_limits} derives the halo limits
and their cross-dataset checks, and Sec.~\ref{sec:uv_translations} translates
those limits onto three UV completions. Sec.~\ref{sec:conclusions} concludes.

%==================================================================
\section{Methodology}
\label{sec:methodology}
%==================================================================
This section sets out the calculation applied throughout the paper: how to
derive, from an effective operator or a UV-complete theory, the exposure
multiplier that observing its scattering signature would require. The gamma-ray datasets to which
the statistic is applied are catalogued in Sec.~\ref{subsec:datasets}.
 
%------------------------------------------------------------------
\subsection{The scattering process}
\label{subsec:kinematics}
%------------------------------------------------------------------
Every result in this work follows from a single $2\to2$ process, elastic
photon--DM scattering,
\begin{equation}
    \gamma(k) + \chi(p) \;\longrightarrow\; \gamma(k\p) + \chi(p\p),
    \label{eq:process}
\end{equation}
the dark-sector counterpart of Compton scattering off an
electron~\cite{KleinNishina1929} (see also
Ref.~\cite[Sec.~5.5]{PeskinSchroeder1995}). Halo DM is
non-relativistic, with velocity dispersion $v_\chi \sim
10^{-3}c$~\cite{Strigari2013DMReview}, so to the accuracy required here, the DM
rest frame coincides with the Galactic frame in which the gamma-ray flux is
measured. All kinematic quantities below are evaluated in that frame.
 
Writing $\omega \equiv E_\gamma$ for the incident photon energy, $\omega\p$ for
the outgoing photon energy, $\theta$ for the scattering angle and $m_\chi$ for
the DM mass, the Mandelstam invariants of Eq.~\eqref{eq:process} are
\begin{equation}
\begin{split}
    s &= m_\chi^{2} + 2 m_\chi \omega, \\[2pt]
    t &= -2 m_\chi\,(\omega - \omega\p),
\end{split}
\label{eq:mandelstam}
\end{equation}
and energy--momentum conservation fixes the outgoing photon energy through the
Compton relation
\begin{equation}
    \omega\p \;=\; \frac{\omega}{1 + (\omega/m_\chi)\,(1-\cos\theta)}\,.
    \label{eq:compton_shift}
\end{equation}
From this kinematic relation, we can already infer the following:
 
First, scattering degrades photons rather than destroying them, since a photon
removed from energy $\omega$ reappears at $\omega\p \le \omega$. The observable
consequence of photon--DM scattering is therefore a distortion of the shape of a gamma-ray spectrum, not an overall deficit in its
normalisation, and the transport problem requires a redistribution kernel in
addition to an attenuation factor (Sec.~\ref{subsec:transport}).
 
Second, the size of the energy shift is controlled by $\omega/m_\chi$. For
$\omega \ll m_\chi$ the photon recoils elastically off an effectively infinitely
heavy target and $\omega\p \to \omega$, so scattering is nearly forward and
nearly energy-conserving; the spectral distortion is correspondingly weak. For
$\omega \gtrsim m_\chi$ the shift becomes $\mathcal{O}(1)$ and the deflection is
large. Gamma-ray energies of $\mathcal{O}(1\text{--}10^{3})~\mathrm{GeV}$
therefore give an $\mathcal{O}(1)$ shift for sub-GeV DM and a progressively
weaker one above it, with the migration closing entirely near
$m_\chi \simeq 1.1~\mathrm{TeV}$, beyond which the signal is carried by angular
deflection out of the region of interest rather than by the spectral shape. A
large shift per scatter does not by itself imply a strong bound, since the
optical depth also carries the column number density $J/m_\chi$ and the mass
dependence of the cross section. For the dipole operators these two effects
cancel in the $\omega \gg m_\chi$ regime, leaving the exclusion flat in mass
(Sec.~\ref{sec:halo_limits}).
 
Third, because the process is a single scatter off a single DM particle, the
attenuation depends on the linear DM column density $\int\rho_\chi\,\dd
l$, the $n=1$ moment of the density profile. This distinguishes it from
annihilation searches, whose signal scales as $\int\rho_\chi^{2}\,\dd l$, and
means that scattering weights the outer halo comparatively more heavily.
 
%------------------------------------------------------------------
\subsection{From interaction to cross section}
\label{subsec:operator_framework}
%------------------------------------------------------------------
The photon--DM interaction enters this work in two ways. In Sec.~\ref{sec:UV_revisit} the mediating
sector is specified explicitly: the SM Higgs with its minimal
dark-Higgs extension, and perturbative quantum gravity. There the amplitude for
Eq.~\eqref{eq:process} is computed directly in the UV-complete theory, and the
free parameters are that theory's own couplings and masses. In Secs.~\ref{sec:eft_operators} and~\ref{sec:halo_limits} no mediator is
assumed. The interaction is instead treated in an effective field theory (EFT) framework,
in which any heavy mediating sector has been integrated out and its effects are
carried by a set of gauge-invariant contact operators built from the DM field,
$\chi$, and the electromagnetic field-strength tensor,
$F_{\mu\nu}$~\cite{KavanaghPanciZiegler2019FaintLight,Arina2021PhotonEFT,Goodman2010EFT},
\begin{equation}
    \mathcal{L} \;=\; \mathcal{L}_{\rm SM} \;+\; \mathcal{L}_\chi
    \;+\; \sum_{i}\frac{c_i}{\Lambda^{\,d_i-4}}\,\mathcal{O}_i \,,
    \label{eq:eft_expansion}
\end{equation}
where $d_i$ is the mass dimension of $\mathcal{O}_i$, $\Lambda$ is the scale at
which the underlying physics is resolved, and $c_i$ is a dimensionless Wilson
coefficient. Throughout we quote bounds on $\Lambda$ at the natural benchmark $c_i = 1$,
since the data constrain only the combination $c_i^{1/n_i}/\Lambda$, with
$n_i = d_i - 4$. Sec.~\ref{sec:uv_translations}
then closes the loop, matching the resulting operator bounds back onto specific
UV completions.
 
The two modes differ only in where the amplitude comes from. The route from an
amplitude to a cross section and everything downstream of it is common to both. That route runs in three steps:
 
\paragraph{Operators.} Each $\mathcal{O}_i$ must be SM gauge invariant, Lorentz
invariant and consistent with the assumed spin and self-conjugation properties
of $\chi$. The complete dimension-5 to dimension-7 basis for scalar, Majorana
and Dirac DM is enumerated in Sec.~\ref{subsec:eft_catalogue}. This step is absent in the UV-complete mode, where the vertex
follows from the Lagrangian of the completion itself.
 
\paragraph{Amplitudes.} For each operator the tree-level amplitude
$\mathcal{M}_i$ for Eq.~\eqref{eq:process} is computed and averaged over initial
and summed over final photon polarisations and DM spins, giving
$\overline{|\mathcal{M}_i|^{2}}(s,t)$. Not every gauge-invariant operator
survives this step with a non-zero result: those whose photon coupling enters
through $\partial^{\nu}F_{\mu\nu}$ vanish on shell, reducing the real-photon
basis to the dipole and Rayleigh families
(Sec.~\ref{subsec:eft_catalogue}).
 
\paragraph{Cross sections.} The differential cross section follows from
$\overline{|\mathcal{M}_i|^{2}}$ by the standard $2\to2$ relation which, for a
massless projectile on a target of mass $m_\chi$, reads
\begin{equation}
    \frac{\dd\sigma_i}{\dd t}
    \;=\;
    \frac{\overline{|\mathcal{M}_i|^{2}}}{16\pi\,\lambda(s,m_\chi^{2},0)}
    \;=\;
    \frac{\overline{|\mathcal{M}_i|^{2}}}{64\pi\,m_\chi^{2}\,\omega^{2}}\,,
    \label{eq:dsigma_dt}
\end{equation}
with $\lambda$ the K{\"a}ll{\'e}n function. Integrating over the kinematically
allowed range of momentum transfer, $t_{\rm min}(\omega,m_\chi) \le t \le 0$
with $|t_{\rm min}| = 4\omega^{2}/(1 + 2\omega/m_\chi)$ from
Eqs.~\eqref{eq:mandelstam}--\eqref{eq:compton_shift}, gives the total cross
section
\begin{equation}
    \sigma_{{\rm tot},i}(\omega, m_\chi;\Lambda)
    \;=\; \int_{t_{\rm min}}^{0}\!\dd t\;\frac{\dd\sigma_i}{\dd t}\,.
    \label{eq:sigma_tot}
\end{equation}

%------------------------------------------------------------------
\subsection{Optical depth and line-of-sight columns}
\label{subsec:optical_depth}
%------------------------------------------------------------------
The probability that a photon of energy $\omega$ is scattered while traversing a
column of DM of length $L$ is the optical depth
\begin{equation}
    \tau_i(\omega)
    \;=\; \frac{J}{m_\chi}\,\sigma_{{\rm tot},i}(\omega, m_\chi;\Lambda),
    \quad
    J \;\equiv\; \int_{0}^{L}\!\rho_\chi(l)\,\dd l ,
    \label{eq:tau_general_revisit}
\end{equation}
where $J$ is the DM column density along the line of sight and
$\rho_\chi/m_\chi$ is the DM number density. Two distinct lines of sight are
considered in this work: a path through the Galactic DM halo, and a
cosmological path through the mean DM density of the universe.
 
Two relevant columns are quoted for reference. The cosmological
baseline~\cite{Planck2018CosmoParams},
\begin{equation}
\begin{split}
    \rho_\chi &= 1.2\times10^{-6}~\mathrm{GeV\,cm^{-3}}, \\
    L &= 1.14\times10^{28}~\mathrm{cm}\;(\approx 3.7~\mathrm{Gpc}),
\end{split}
\label{eq:cosmo_baseline}
\end{equation}
gives $J_{\rm cosmo} = 1.37\times10^{22}~\mathrm{GeV\,cm^{-2}}$. The Galactic
baseline uses the local
DM density and the Sun's Galactocentric
distance~\cite{Read2014LocalDensity,deSalasWidmark2021LocalDensity,%
GRAVITY2019Distance},
\begin{equation}
\begin{split}
    \rho_\chi &= 4\times10^{-1}~\mathrm{GeV\,cm^{-3}}, \\
    L &= 2.62\times10^{22}~\mathrm{cm}\;(\approx 8.5~\mathrm{kpc}),
\end{split}
\label{eq:galactic_baseline}
\end{equation}
and gives $J_{\rm gc} = 1.05\times10^{22}~\mathrm{GeV\,cm^{-2}}$. Neither is
used to set a bound. Both hold the density fixed along the path, and the
cosmological baseline is unredshifted and neglects the growth of structure. They are
retained because they are the baselines of
Ref.~\cite{acar2026darkmatterredblue}, against which
Sec.~\ref{sec:UV_revisit} is benchmarked on the same footing. $J_{\rm cosmo}$ is
used once more, as the column of the isotropic diffuse cross-check of
Sec.~\ref{subsec:cross_dataset}, where a constant-density treatment is
appropriate to an isotropic source. Fixing $J_{\rm gc}$ also fixes the scale against which the
improvement of the present analysis can be read: the emissivity-weighted,
region-of-interest-averaged (ROI-averaged) column $J_{\rm ROI}$ actually used
for the bounds of Secs.~\ref{sec:eft_operators} and~\ref{sec:halo_limits},
defined in Sec.~\ref{subsec:halo_dataset}, is a factor $\approx\!4.6$ larger
than $J_{\rm gc}$ because it integrates a Navarro--Frenk--White (NFW)
profile~\cite{NFW1996} over the ROI rather than assuming a
uniform density over a fixed path length.
 
%------------------------------------------------------------------
%------------------------------------------------------------------
\subsection{Spectral transport and the test statistic}
\label{subsec:transport}
%------------------------------------------------------------------
Because scattering softens photons rather than removing them, an intrinsic
source spectrum $\Phi_{\rm src}(E_\gamma)$ is transported into a predicted
observed spectrum by an attenuation term and a scattering term,
\begin{equation}
\begin{split}
    \Phi_{\rm pred}(E_\gamma) = e^{-\tau(E_\gamma)}\,\Phi_{\rm src}(E_\gamma)&\\
    + \int\!\dd E\p\,K(E_\gamma\,|\,E\p)\,\tau(E\p)&\,e^{-\tau(E\p)}\,
    \Phi_{\rm src}(E\p),
\end{split}
\label{eq:atten+reshape}
\end{equation}
where the first term is survival (pure attenuation), the second is a single
in-scatter with the Poisson weight $\tau e^{-\tau}$, and the redistribution
kernel $K(E_\gamma\,|\,E\p)$ is the normalised probability that a photon
scattered at energy $E\p$ emerges at $E_\gamma$. Its explicit construction and normalisation are
\begin{equation}
\begin{split}
    K(E\,|\,E\p) = &\frac{\Theta(E\p - E)}{\sigma_{\rm tot}(E\p)}\, 
    \frac{\dd\sigma}{\dd E}(E\p\!\to\!E), \\
    \int_{0}^{E\p}&\!\dd E\;K(E\,|\,E\p) = 1,
\end{split}
\label{eq:K_app}
\end{equation}
so it is normalised over the outgoing energy at fixed $E\p$. The step function makes
$K$ upper-triangular, since Compton kinematics only degrades the energy,
$E \leq E\p$.

A scatter changes the photon's direction as well as its energy, and the two
observables account for the deflection differently. In the attenuation
observable, deflection enters as part of the removal criterion: a photon deflected out of the
region of interest is removed, weighted by the template-computed recovery
fraction $w(\theta)$ entering Eq.~\eqref{eq:removal_two_channel}. In the
reshaping observable, an in-scattered photon is retained only to the extent
that it remains within the ROI, so the kernel actually applied
carries the same weight,
$K_{w}(E\,|\,E\p) = w\bigl(\theta(E\p\!\to\!E)\bigr)\,K(E\,|\,E\p)$, with the
deflection angle fixed by the energy pair through the Compton relation. Its
column sums obey $\int\dd E\,K_{w} \leq 1$, with equality only for
$w \equiv 1$; photons deflected beyond the region of interest leave the
bookkeeping, as they should. Exact number conservation holds for an isotropic
source against a uniform scatterer, where a deflected photon leaves one line
of sight only to enter an equivalent one, the situation of the isotropic
cross-check in Sec.~\ref{subsec:cross_dataset}.

The pure-attenuation observable retains only the survival term,
$\Phi_{\rm pred} = e^{-\tau}\Phi_{\rm src}$, while the full reshaping
observable retains both. On the discrete halo-posterior energy grid, the
integral in Eq.~\eqref{eq:atten+reshape} becomes the matrix sum used in
Sec.~\ref{subsec:halo_results}.

In place of a single $\tau$-threshold detectability criterion we adopt a
continuous, data-driven metric: the $\chi^{2}$ distance between the measured
spectrum, $\Phi_{\rm data}$, and the transported prediction of
Eq.~\eqref{eq:atten+reshape},
\begin{equation}
    \Delta\chi^{2}(m_\chi, \Lambda)
    \;=\;
    \sum_{i}\,
    \frac{\bigl[\Phi_{\rm data}(E_i) - A\,\Phi_{\rm pred}(E_i\,;\,m_\chi,\Lambda)\bigr]^{2}}
         {\sigma_i^{2}},
    \label{eq:delta_chi2_general}
\end{equation}
profiled over the source normalisation $A$ and referenced to the minimum over the scanned grid, $\Delta \chi^2 \equiv \chi^2 - \chi^2_{\text{min}}$, with $\sigma_i$ the bin-level statistical uncertainty of the measurement. Because $\Phi_{\text{src}} = \Phi_{\text{data}}$, the minimum sits at $\tau = 0$ for the halo fit, so $\Delta \chi ^2$ and $\chi^2$ coincide there. Profiling $A$ removes the
energy-independent component of the attenuation, which is degenerate with the
unknown absolute normalisation of the source, so the statistic tests the shape distortion alone.

The intrinsic template $\Phi_{\rm src}$ entering Eq.~\eqref{eq:atten+reshape}
must be specified separately from the data, and cannot be left free. With
$\Phi_{\rm src}$ unconstrained, the transport of Eq.~\eqref{eq:atten+reshape}
can be inverted exactly, being triangular with positive diagonal $e^{-\tau}$ on
the discrete energy grid, so a source spectrum exists that reproduces the
measurement for every operator at every $\Lambda$: an unconstrained fit
excludes nothing, since any amount of scattering can be absorbed into a
sufficiently contrived intrinsic spectrum. The question the data can
answer is therefore not whether some source spectrum survives scattering, but
how large a spectral distortion a given operator would imprint relative to the
measured uncertainties. Our default choice is
$\Phi_{\rm src} = \Phi_{\rm data}$, which poses exactly that question and is
self-consistent in the regime of interest: since
$\Phi_{\rm data} = \Phi_{\rm src} + \mathcal{O}(\tau)$, using the measurement as
its own template is correct to leading order, with the error entering at
$\mathcal{O}(\tau^{2})$ and $\tau \lesssim 3\times10^{-2}$ on the attenuation
contours. The reshaping arm reaches larger optical depths and carries its own
validity discussion in Sec.~\ref{subsec:halo_results}.
 
Because the same spectrum appears on both sides of
Eq.~\eqref{eq:delta_chi2_general}, $\chi^{2}$ vanishes identically as
$\tau \to 0$, so $\Delta\chi^{2}(m_\chi,\Lambda)$ is the spectral distortion a
given operator would imprint, measured in units of the bin uncertainties.

We take $\Delta\chi^{2} = 4.61$ as the reference at which that distortion is
resolved, the $90\%$ confidence-level value for a scan over the two parameters
$(m_\chi,\Lambda)$~\cite{ParticleDataGroup2024}; the joint two-parameter
convention is conservative relative to the one-parameter raster
($\Delta\chi^{2} = 2.71$) common in indirect-detection analyses.

The regularity conditions behind the asymptotic $\chi^{2}$ calibration are not
satisfied here, so we calibrate the threshold directly rather than assume it. The
optical depth is bounded from below, $\tau \geq 0$, so the null lies on the
boundary of the parameter space and the asymptotic distribution is a mixture
rather than a pure $\chi^{2}$. Separately, $m_\chi$ is not identified under the
null, since the model is independent of the scatterer mass when $\tau = 0$; Ref.~\cite{Davies1987} treats this case. The two effects act in opposite
directions and do not cancel in any controlled way.

We therefore calibrate against pseudo-experiments drawn from the per-bin
posterior chains of Ref.~\cite{StenhouseGhagDeppisch2026Totani}, generated at the
signal hypothesis so that coverage is defined. Across twelve points on the
exclusion contour, spanning three operators and four decades in DM mass,
the calibrated $90\%$ quantile of the test statistic lies between $3.51$ and
$4.02$, with median $3.84$, and shows no trend with mass. The adopted threshold
$\Delta\chi^{2} = 4.61$ therefore over-covers slightly, delivering $92$--$95\%$
coverage against the nominal $90\%$. The direction is the one the boundary
effect predicts, and the contours quoted below are correspondingly conservative.

We retain $4.61$ in preference to the calibrated $3.84$ for a reason independent
of coverage. Because the statistic is referenced to the minimum over the scanned
grid, part of its budget is spent on the scan before any signal is present: under
the null the scan minimum improves on the $\tau = 0$ point by
$\Delta\chi^{2} = 2.05$ at the $90$th percentile, so a threshold of $3.84$ would
exclude the no-scattering hypothesis itself in $2.7\%$ of pseudo-experiments,
against $1.5\%$ at $4.61$. The median expected contour evaluated at the
calibrated threshold agrees with the contour quoted here to better than a tenth
of one grid step in $\Lambda$.

The contours below are sensitivity estimates in the frequentist sense just
described: the threshold is calibrated against pseudo-experiments and the
coverage is correct, so each contour marks where a $90\%$~CL exclusion would
lie for the operator, the halo model and the assumed Wilson coefficient.
Because the source template is the measured spectrum itself, the statistic
carries no fluctuations and the contours are median-expected exclusions rather
than observed ones, which is why we refer to them throughout as sensitivity.
They do not constrain a viable DM model in any case. As
Secs.~\ref{subsec:halo_results} and~\ref{sec:uv_translations} show, the
regions they cover lie below the EFT-validity threshold and admit no
perturbative UV completion. Where we write that a contour excludes a scale,
the statement is about effective operator parameter space.

A photon-counting measurement has $\sigma_i \propto
1/\sqrt{\smash[b]{\text{exposure}}}$, so at fixed spectral shape the statistic
of Eq.~\eqref{eq:delta_chi2_general} scales linearly with exposure. A point in
the $(m_\chi,\Lambda)$ plane that currently returns
$\Delta\chi^{2}_{\rm current}$ would therefore reach that reference under an
exposure larger by the factor
\begin{equation}
    f_{\rm required}(m_\chi, \Lambda)
    \;=\;
    \frac{4.61}{\Delta\chi^{2}_{\rm current}(m_\chi, \Lambda)}.
    \label{eq:f_required}
\end{equation}
We report $f_{\rm required}$ rather than a binary detected/not-detected verdict
for three reasons: it is continuous, so it distinguishes between regions that
are marginally and hopelessly out of reach; it depends on the reference value
only through an overall rescaling, so the ranking of the parameter space is
independent of that choice; and it states directly how much additional data a
given operator would need, so the same map can be read against the exposure of
any future instrument. The framework itself extends further. Given per-bin
fluxes with uncertainties and a line-of-sight column density, the pipeline
recomputes the sensitivity contours for any instrument without modification, and only the reading of
$f_{\rm required}$ as a pure exposure multiplier is tied to a fixed energy
band, since a change of band changes the spectral shape the statistic is built
on (Sec.~\ref{subsec:halo_results}). 
 
%------------------------------------------------------------------
\subsection{Datasets}
\label{subsec:datasets}
%------------------------------------------------------------------
Five datasets or reference spectra are used, and all five are characterised
here. Two carry the main results. Of the remaining three, one is the theory-only
input of Sec.~\ref{sec:UV_revisit} and the other two enter only in
Sec.~\ref{subsec:cross_dataset}.
 
\paragraph{Halo posterior.}
The primary spectral dataset is the pixel-level Markov Chain Monte Carlo (MCMC)
halo posterior of Ref.~\cite{StenhouseGhagDeppisch2026Totani}, extracted from 17
years of \textit{Fermi}--LAT Pass~8 ULTRACLEAN photons~\cite{Atwood2009FermiLAT}
over the sky region of Ref.~\cite{totani202520gevhalolikeexcess}. The data
reduction, the point-source and diffuse-background templates, the
energy-dependent point-spread function forward folding, and the resulting halo
posteriors under the NFW profiles $\rho^{2}$ and $\rho^{2.5}$ are all defined
and characterised in Ref.~\cite{StenhouseGhagDeppisch2026Totani}.

\paragraph{IGRB.}
The independent cross-check dataset is the isotropic diffuse gamma-ray
background (IGRB) measured by the \textit{Fermi}--LAT collaboration in
Ref.~\cite{Ackermann2015IGRB} (Foreground Model~A, 26 bins from
$100~\mathrm{MeV}$ to $820~\mathrm{GeV}$). Its sky coverage, foreground
modelling and reduction pipeline are independent of the halo posterior, making it a suitable systematic cross-check.
 
\paragraph{Perturbative UV completions (theory only).}
The $m_\chi$--$\tau$ analysis of Sec.~\ref{sec:UV_revisit} and
Fig.~\ref{fig:mchi_vs_tau} is a purely theoretical calculation, unconnected to
any gamma-ray measurement. Its curves are the maximum optical depth achievable
within the minimal dark-Higgs extension (saturating perturbativity and the Large Hadron Collider (LHC) signal-strength ceiling), and within perturbative quantum
gravity, evaluated at $E_\gamma = 175~\mathrm{GeV}$.
 
\paragraph{Annihilation templates.}
The two-component variants of Sec.~\ref{subsec:cross_dataset} replace
$\Phi_{\rm src}$ with the prompt gamma-ray yield from DM annihilation, tabulated
in Ref.~\cite{Cirelli2011PPPC}. These variants assume a two-component dark
sector, in which the particle that annihilates is a distinct species from the
light particle that scatters. Where the two must be distinguished we write
$m_\chi^{\rm ann}$ for the annihilator mass, held fixed at the best-fit values
of Ref.~\cite{StenhouseGhagDeppisch2026Totani}, and $m_\chi^{\rm scat}$ for the
scatterer mass, which remains the only free mass in the fit. Elsewhere in this
work a single DM species is assumed and its mass is written $m_\chi$ without
adornment.
 
\paragraph{dSph SEDs~\cite{McDaniel2024DsphLegacy} and density profiles~\cite{GeringerSameth2015Dfactors}.}
These enter only the feasibility discussion of Sec.~\ref{subsec:cross_dataset}.
No bound is derived from this dataset.

%==================================================================
\section{UV-Complete Scattering Revisited}
\label{sec:UV_revisit}
%==================================================================

We now apply the framework of Sec.~\ref{sec:methodology} to the two UV
completions considered in Ref.~\cite{acar2026darkmatterredblue}: the SM Higgs
portal with its minimal dark-Higgs extension, and perturbative quantum gravity.
For each we ask how large the coupling must be for $\tau(E_\gamma)$ to reach a
magnitude which a current instrument could resolve, and whether that coupling lies inside
the perturbatively consistent regime. As the scale for ``resolvable'' we take
$\tau \approx 2.6\times10^{-2}$, the inverse-variance-combined statistical
uncertainty of the \textit{Fermi}--LAT halo posterior of
Sec.~\ref{sec:halo_limits} ($2.58\%$ across the eight retained bins), drawn as a
horizontal dotted line in
Fig.~\ref{fig:mchi_vs_tau}. It serves for orientation only. Every quantitative
statement in this work rests on Eqs.~\eqref{eq:delta_chi2_general}
and~\eqref{eq:f_required}. Beyond that single reference value the calculations
of this section are purely theoretical, with no spectral dataset fitted, so
their conclusions are independent of the choice of gamma-ray data.

\subsection{Weakly Interacting Case}
\label{subsec:weak_case}

In the simplest realisation considered in
Ref.~\cite{acar2026darkmatterredblue}, the DM mass is generated entirely by
electroweak symmetry breaking through a direct coupling to the SM Higgs.
Written after symmetry breaking, in terms of the physical scalar, $h$, the
interaction is $\mathcal{L} \supset -y_\chi\,\bar{\chi}\chi\,(v+h)$ with
\begin{equation}
    y_\chi = \frac{m_\chi}{v},
    \qquad
    v = 246~\mathrm{GeV}.
    \label{eq:y_chi_direct}
\end{equation}

The status of this construction bounds how far its conclusions can be pushed.
For a $\chi$ that is a singlet
under the SM gauge group, $\bar{\chi}\chi H$ is not gauge invariant: a
renormalisable Yukawa with the Higgs doublet requires $\chi$ to exist within an
electroweak multiplet, in which case it also couples to the $W$ and $Z$ and
inherits the associated collider and direct-detection constraints. The SM-gauge-invariant alternative for a singlet is the dimension-5 operator
$(c/\Lambda)\,\bar{\chi}\chi\,H^{\dagger}H$, which, after symmetry breaking,
yields both a DM mass and an $h\bar{\chi}\chi$ coupling of the form assumed
here, but which is itself an effective operator rather than a UV completion.
The direct portal is therefore best read as a low-energy parameterisation, and
this subsection as an assessment of how far that parameterisation can be
taken.

The differential cross section for $\gamma\chi \to \gamma\chi$ through this
coupling is~\cite{acar2026darkmatterredblue}
\begin{equation}
\begin{split}
    \frac{\dd\sigma}{\dd\Omega}
    &= y_{\chi}^{\,2} \Big( \frac{\alpha}{\pi v} \Big)^2\, \frac{3t^2}{8}\,
       \frac{2m_\chi^2 - t/2}{(t-m_H^2)^2 + m_H^2\Gamma_H^2} \\
    &\quad \times \frac{\bigl|I_W(\tau_W)
       + \sum_f N_c\,Q_f^2\,I_f(\tau_f)\bigr|^2}{64\pi^2\,s}\,,
\end{split}
\label{eq:dsdo_revisit}
\end{equation}
where $I_W(\tau_W)$ and $I_f(\tau_f)$ are the spin-1 and spin-$\tfrac{1}{2}$
Passarino--Veltman form factors~\cite{PassarinoVeltman1979}, with
\begin{equation*}
    \tau_W = \frac{4 m_W^2}{|t|}, \qquad \tau_f = \frac{4 m_f^2}{|t|},
\end{equation*}
and $s$, $t$ the Mandelstam invariants of Eq.~\eqref{eq:mandelstam}, evaluated
in the DM rest frame; $|t| = -t$ is the spacelike photon momentum transfer. The
sum runs over the charged fermions $f$ in the loop and is top-dominated. The
overall factor $y_\chi^2$ enters the optical depth as $\tau \propto y_\chi^2$
at fixed loop kinematics.

A standard criterion for perturbative validity is that the relevant Yukawa
coupling satisfy
\begin{equation}
    y_\chi \lesssim \sqrt{4\pi} \approx 3.54,
    \label{eq:pert_criterion}
\end{equation}
beyond which the one-loop correction to the
$\gamma\chi\to\gamma\chi$ amplitude becomes comparable to the tree-level
result, and the loop expansion ceases to be controlled. Combining Eq.~\eqref{eq:y_chi_direct} with
Eq.~\eqref{eq:pert_criterion} yields a perturbativity ceiling on the DM mass,
\begin{equation}
    m_\chi \;\lesssim\; \sqrt{4\pi}\,v \;\approx\; 870~\mathrm{GeV},
    \label{eq:mchi3_pert_ceiling}
\end{equation}
so the direct portal is consistent at one loop only for
$m_\chi \lesssim 1~\mathrm{TeV}$. The benchmark $m_\chi = 1~\mathrm{TeV}$ used
throughout Ref.~\cite{acar2026darkmatterredblue} sits at that boundary, and
extrapolation to heavier DM, natural given
$\sigma \propto m_\chi^{2}$, is external to the regime in which the loop
calculation is valid.

Two independent considerations therefore point the same way: the construction
is not a UV completion, and even taken at face value it cannot be extrapolated
past the electroweak scale. To probe heavier DM in a controlled way, the direct
portal must be replaced by a construction that decouples the DM mass from the
SM vacuum expectation value.

%------------------------------------------------------------------
\subsection{Minimal Dark-Higgs Extension}
\label{subsec:dark_higgs}
%------------------------------------------------------------------

The simplest extension that achieves this decoupling is a minimal dark-Higgs
portal~\cite{PattWilczek2006}. The SM scalar sector is augmented by a single
real scalar $H\p$, a singlet under all SM gauge groups, which acquires a vacuum
expectation value $v\p$ through a hidden symmetry-breaking pattern in the dark
sector. The DM couples to it through a dark-sector Yukawa,
\begin{equation}
    \mathcal{L}_{\rm dark} \supset -y_\chi\,\bar{\chi}\chi\,H\p,
    \qquad
    y_\chi = \frac{m_\chi}{v\p},
    \label{eq:y_chi_dark}
\end{equation}
which, unlike its SM counterpart in Sec.~\ref{subsec:weak_case}, is SM gauge
invariant as written, precisely because $H\p$ is a singlet. This is therefore a
genuine renormalisable completion rather than a low-energy parameterisation,
and it is the Higgs portal version we carry forward.

The two scalar sectors communicate through the renormalisable portal operator
$\lambda_{HP}\,(H^{\dagger}H)\,H^{\prime 2}$. 
After electroweak and dark symmetry
breaking, this mixes the two neutral $CP$-even states, giving mass eigenstates
$(h, h\p)$ related to the gauge eigenstates by an angle $\theta$,
\begin{align}
    h   &= H\cos\theta + H\p\sin\theta, \\
    h\p &= -H\sin\theta + H\p\cos\theta.
\end{align}
$H$ denotes the neutral $CP$-even component of the SM doublet. The
light eigenstate $h$ is identified with the 125~GeV scalar observed at the LHC,
and the heavy eigenstate $h\p$ is integrated out at the energies of interest.
The effective $\bar{\chi}\chi h$ coupling inherited by the SM-like Higgs is
\begin{equation}
    y_{\chi H} \;\equiv\; y_\chi\,\sin\theta \;=\; \frac{m_\chi}{v\p}\,\sin\theta.
    \label{eq:y_chi_H}
\end{equation}

Two features follow immediately. First, $y_\chi$ is decoupled from the SM
vacuum expectation value: taking $v\p \gtrsim m_\chi/\sqrt{4\pi}$ keeps the
dark-sector Yukawa perturbative for arbitrarily heavy DM. Second, $\sin\theta$
is directly constrained by the measured signal strength of the 125~GeV Higgs.
Combined fits to ATLAS and CMS Run-2 data restrict the mixing
to~\cite{ATLASHiggs2022, CMSHiggs2022}
\begin{equation}
    \sin^2\theta \;\lesssim\; 0.11
    \Longrightarrow
    \sin\theta \;\lesssim\; 0.33
    \ (95\%~\mathrm{CL}),
    \label{eq:mixing_bound}
\end{equation}
which we adopt as the nominal upper bound in what follows. Both constraints act
on the dimensionless coupling $y_{\chi H}$ of Eq.~\eqref{eq:y_chi_H}, which is
the quantity the dark sector controls.

The Higgs-mediated $\gamma\chi\to\gamma\chi$ amplitude,
Eq.~\eqref{eq:dsdo_revisit}, carries through unchanged in this extended
scenario under the substitution $y_\chi \to y_{\chi H}$ wherever the
DM--SM-Higgs coupling appears. The cross section is then controlled by the
effective vertex coefficient
\begin{equation}
    \mathcal{C}_{\rm eff}
    \;\equiv\;
    \frac{\alpha}{\pi v}\,y_{\chi H}
    \;=\;
    \frac{\alpha}{\pi v}\,\frac{m_\chi}{v\p}\,\sin\theta ,
    \label{eq:y_eff_def}
\end{equation}
of mass dimension $-1$, in which
$\alpha/(\pi v) = 9.4\times10^{-6}~\mathrm{GeV^{-1}}$ is a fixed SM loop
prefactor and $y_{\chi H}$ carries all model-dependent dark-sector physics.
%---
%------------------------------------------------------------------
\subsection{Maximum perturbative signal: the $m_\chi$--$\tau$ plane}
\label{subsec:mchi_tau_plot}
%------------------------------------------------------------------

\begin{figure}[htbp]
    \centering
\includegraphics[width=\linewidth]{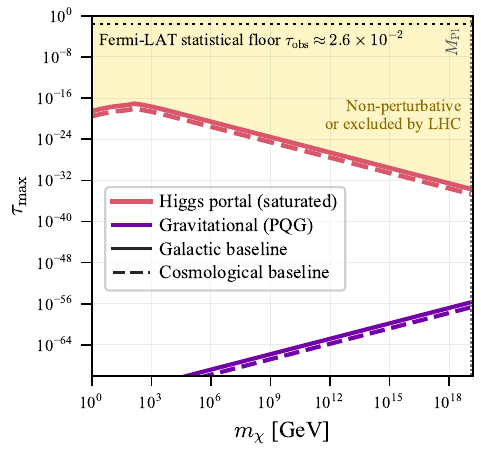}
    \caption{
        Maximum optical depths $\tau_{\rm max}(m_\chi)$ achievable within the
        minimal dark-Higgs extension and the gravitationally interacting case,
        evaluated at $E_\gamma = 175~\mathrm{GeV}$, near the top-quark loop
        threshold where the $H\to\gamma\gamma$ form factor peaks. Colour
        distinguishes the two cases as labelled, and within each the solid
        curve is the galactic baseline and the dashed curve the cosmological
        one. The horizontal dotted line at $\tau \approx 2.6\times10^{-2}$
        marks an approximate detection target equal to the
        inverse-variance-combined statistical uncertainty of the
        \textit{Fermi}--LAT Galactic-centre halo posterior. The dotted vertical
        guide marks the Planck mass
        $M_{\rm Pl} = 1.22\times10^{19}~\mathrm{GeV}$, the heaviest sensible
        point-particle scatterer. The pink curve simultaneously saturates the
        dark-sector perturbativity bound $y_\chi = \sqrt{4\pi}$ and the LHC
        Higgs-mixing bound $\sin\theta = 0.33$. The shaded band above is
        accessible only by violating one or both of these constraints.
    }
    \label{fig:mchi_vs_tau}
\end{figure}

We now combine the perturbativity ceiling on $y_\chi$ with the LHC mixing-angle bound on $\sin\theta$ to extract the largest optical depth achievable within the perturbatively consistent dark-Higgs extension. Saturating both constraints,
\begin{equation}
    y_\chi^{\rm max} = \sqrt{4\pi},
    \qquad
    \sin\theta^{\rm max} = 0.33,
    \label{eq:saturated_couplings}
\end{equation}
the maximum effective vertex coefficient is
\begin{equation}
\begin{split}
    \mathcal{C}_{\rm eff}^{\rm max}
    \;&=\;
    \frac{\alpha}{\pi v}\,\sqrt{4\pi}\,\sin\theta^{\rm max}
    \;\approx\;
    \frac{\alpha}{\pi v}\cdot 1.2 \\
    \;&\approx\;
    1.1\times10^{-5}~\mathrm{GeV}^{-1}.
    \label{eq:Ceff_max}
\end{split}
\end{equation}
In this saturated regime the dimensionless coupling
$y_{\chi H}^{\rm max} = \sqrt{4\pi}\,\sin\theta^{\rm max} \approx 1.2$, and
hence $\mathcal{C}_{\rm eff}^{\rm max}$, are independent of $m_\chi$: the two
factors of $m_\chi$ that would otherwise enter through the dark-sector Yukawa
($y_\chi = m_\chi/v\p$) are absorbed into the choice
$v\p = m_\chi/\sqrt{4\pi}$ that saturates perturbativity. The entire mass
dependence of the maximum signal is therefore carried by the kinematic
structure of the loop and by the column-density factor $J/m_\chi$ in
Eq.~\eqref{eq:tau_general_revisit}.

Fig.~\ref{fig:mchi_vs_tau} shows the resulting $\tau_{\rm max}(m_\chi)$,
obtained by inserting Eqs.~\eqref{eq:Ceff_max}
and baselines~\eqref{eq:cosmo_baseline}/\eqref{eq:galactic_baseline} into
Eq.~\eqref{eq:tau_general_revisit} and integrating
Eq.~\eqref{eq:dsdo_revisit} numerically over the scattering angle at
$E_\gamma = 175~\mathrm{GeV}$, chosen to sit near the top-quark loop threshold
($4m_{\rm top}^{2}/|t| \sim 1$ in backscattering, $\theta = \pi$, at
$m_\chi \gg E_\gamma$) where the $H\to\gamma\gamma$ form factor is maximised.
The curve rises through the loop region, where $\sigma \propto m_\chi^{2}$
while the number density falls as $m_\chi^{-1}$, giving a net
$\sigma/m_\chi \propto m_\chi$. It peaks near $m_\chi \sim 1~\mathrm{TeV}$,
where the $W$ and top form factors are most efficient for the chosen photon energy,
$E_\gamma = 175~\mathrm{GeV}$. Above that mass the form factors and the cross section both become independent of $\mchi$, and the fall is carried entirely by the number-density factor $J/\mchi$. Across the full displayed range
$\tau_{\rm max}$ remains many orders of magnitude below the
$\tau \approx 2.6\times10^{-2}$ reference line.

Translated through Eq.~\eqref{eq:f_required}, detecting the saturated
dark-Higgs extension at $90\%$~CL would require an exposure boost
$f_{\rm required} \gg 10^{20}$ throughout the probed mass range. The dark-Higgs extension
removes the strict $m_\chi \lesssim \sqrt{4\pi}\,v$ ceiling of the direct
portal and decouples the dark Yukawa from the SM vacuum expectation value, yet
measurable spectral attenuation still demands effective couplings far outside
the perturbative regime in which the loop calculation is internally consistent.
%------------------------------------------------------------------
\subsection{The Gravitational Case}
\label{subsec:grav_revisit}
%------------------------------------------------------------------

The second UV-complete scenario considered in Ref.~\cite{acar2026darkmatterredblue} is photon--DM scattering mediated by a single graviton exchange within perturbative quantum gravity. The cross section is proportional to $G_N^2 \propto M_{\rm Pl}^{-4}$, so that in the heavy-DM limit
\begin{equation}
    \sigma_{\rm tot}^{\rm grav} \;\propto\; \frac{m_\chi^2}{M_{\rm Pl}^4}
    \quad\Longrightarrow\quad
    \tau_{\rm grav} \;\propto\; \frac{J\,m_\chi}{M_{\rm Pl}^4},
    \label{eq:tau_grav_scaling}
\end{equation}
which is independent of $E_\gamma$ in the regime $m_\chi \gg E_\gamma$.
Crucially, unlike the Higgs-portal case, the gravitational amplitude contains no free coupling: $G_N$ is fixed by measurement, and the only model-dependent parameter is $m_\chi$ itself. 
Perturbative quantum gravity is internally consistent as a low-energy effective theory provided $\sqrt{s} \ll M_{\rm Pl}$, a condition trivially satisfied for any DM mass below the Planck scale and any photon energy of astrophysical interest.

Ref.~\cite{acar2026darkmatterredblue} obtains the optical depth as $\tau_{\rm grav} \lesssim 10^{-57}$ even at $m_\chi = M_{\rm Pl}$ along a cosmological baseline, some 55 orders of magnitude below the inverse-variance noise floor of the \textit{Fermi}--LAT halo posterior. That result therefore requires no further mathematical refinement. The calculation is already performed within its regime of validity and contains no adjustable couplings to optimise over. The graviton-exchange differential cross section is forward divergent, so the $\tau_{\text{grav}}$ curve of Fig.~\ref{fig:mchi_vs_tau} is evaluated with the same removal criterion as Eq.~\eqref{eq:removal_criterion}; the conclusion is insensitive to that choice at the fifty-order-of-magnitude level of the shortfall. The gravitational channel is fundamentally limited by the smallness of $G_N$, not by perturbativity considerations, and remains many orders of magnitude beyond observational reach in any astrophysical environment considered, as shown in Fig.~\ref{fig:mchi_vs_tau}.

%==================================================================
\section{EFT Operator Scan}
\label{sec:eft_operators}
%==================================================================

We now reformulate the analysis of Sec.~\ref{sec:UV_revisit} within a
model-independent EFT framework. That section showed two specific mediator
structures, a Higgs portal whose coupling
is capped by perturbativity and the LHC Higgs-mixing bound, and a gravitational
channel suppressed by $G_N^{2}$, to be unobservable. However, a null result on two completions does not preclude the status of photon--DM scattering as an observation channel:
whether
\emph{any} consistent local operator can produce observable scattering at
scales reachable by present or near-future instruments remains open.
We therefore integrate out the UV physics
responsible for the photon--DM coupling at a scale $\Lambda$, leaving a tower
of higher-dimensional local operators suppressed by inverse powers of
$\Lambda$. For each operator, we compute the optical depth along a fixed
astrophysical line of sight and scan the resulting $(\Lambda, m_\chi)$ plane
for regions in which the predicted signal is both observationally relevant and
consistent with the EFT validity criteria.

%------------------------------------------------------------------
\subsection{Operator Catalogue}
\label{subsec:eft_catalogue}
%------------------------------------------------------------------

The structure of the operator basis depends on the spin and self-conjugacy properties of the DM candidate.  
We consider three cases: a real scalar $\phi$, a Majorana fermion $\chi$ (self-conjugate), and a Dirac fermion $\psi$ (complex).  
In each case we restrict to $U(1)_{\rm em}$-invariant, Lorentz-invariant,
Hermitian operators of dimension $5 \leq d \leq 7$ built from the DM bilinear, the photon field strength, $F_{\mu\nu}$, its dual,
$\tilde{F}^{\mu\nu} = \tfrac{1}{2}\epsilon^{\mu\nu\rho\sigma}F_{\rho\sigma}$, or
their derivatives, normalised by the appropriate power of $\Lambda$. These are
operators of the electroweak-broken phase. Above the electroweak vacuum expectation value $v = 246~\mathrm{GeV}$, the
corresponding basis is built from $B_{\mu\nu}$ and $W^{3}_{\mu\nu}$, so an
operator written here in terms of $F_{\mu\nu}$ descends from a combination of
the two~\cite{Arina2021PhotonEFT}.

Refs.~\cite{PospelovterVeldhuis2000,Arina2021PhotonEFT} catalogue the
electromagnetic moments available to neutral dark matter, the charge radius, the
magnetic and electric dipoles and the anapole, together with their role in
detection. Of these one-photon moments only the anapole survives for a Majorana
fermion~\cite{HoScherrer2013,KoppMichaelsSmirnov2014}, because
$\bar\chi\gamma^{\mu}\chi$ and $\bar\chi\sigma^{\mu\nu}\chi$ vanish identically
for a self-conjugate field, whereas the diagonal
dipoles~\cite{Sigurdson2004DipoleDM,FortinTait2012} require a Dirac state.
Two-photon Rayleigh couplings are open to both cases and are given in
Eq.~\eqref{eq:O_rayleigh_majorana}.

\paragraph{Scalar DM.} For a real scalar stabilised by a $Z_2$ symmetry
$\phi \to -\phi$, the only parity-even operator coupling $\phi$ to two photons
at $d \leq 7$ is the scalar Rayleigh operator~\cite{Barducci2025ScalarRayleigh}
\begin{equation}
    \mathcal{O}_{\rm R}^{\phi}
    \;=\;
    \frac{c_\phi}{\Lambda^{2}}\,\phi^{2}\,F_{\mu\nu}F^{\mu\nu}.
    \label{eq:O_rayleigh_scalar}
\end{equation}
This is a dimension-6 operator, with cross section scaling as
$\sigma \propto E_\gamma^4/(\Lambda^4 m_\phi^2)$. The $Z_2$ assumption removes
the lower-dimensional $\phi F_{\mu\nu}F^{\mu\nu}$, which carries an odd power of
$\phi$ and would otherwise appear at $d=5$. Some such symmetry is required in
any case to render $\phi$ stable, and abandoning it opens no new channel here.
A single $\phi$ leg cannot close an elastic $\gamma\phi \to \gamma\phi$
amplitude, so $\phi F_{\mu\nu}F^{\mu\nu}$ contributes only at one loop, and the
operator instead mediates $\phi \to \gamma\gamma$ at a rate
$\Gamma = c_\phi^{2} m_\phi^{3}/(4\pi\Lambda^{2})$. A lifetime consistent with
gamma-ray decay bounds~\cite{Slatyer2021IndirectReview} then requires
$\Lambda \gtrsim 3\times10^{25}~\mathrm{GeV}$ at $m_\phi = 1~\mathrm{GeV}$,
rising as $m_\phi^{3/2}$, six orders of magnitude above the Planck scale and far
outside the range probed here. 
The parity-odd
$\phi F_{\mu\nu}\tilde{F}^{\mu\nu}$ is excluded by the same argument. Following the scalar-Rayleigh convention of Ref.~\cite{Barducci2025ScalarRayleigh}, no factor of $1/4$ is included here. This differs from the fermionic Rayleigh operators of Eq.~\eqref{eq:O_rayleigh_majorana}, where the $1/4$ of Ref.~\cite{WeinerYavin2012RayDM} is retained. (Quoting the scalar limit in the Weiner--Yavin convention instead rescales $c_\phi \to c_\phi/4$ and halves the quoted $\Lambda$. The scalar and fermionic panels can therefore be compared at a common convention with this factor.) 

The parity-odd partner of Eq.~\eqref{eq:O_rayleigh_scalar}, $\phi^{2}F_{\mu\nu}\tilde{F}^{\mu\nu}$, exists at the same dimension. 
It is not scanned separately: for a scalar two-photon coupling the $CP$-even and $CP$-odd vertices have identical polarisation-summed matrix elements, so its unpolarised elastic cross section coincides with that of $\phi^{2}F_{\mu\nu}F^{\mu\nu}$. The two differ only in the linear-polarisation correlations of the scattered photons. 

\paragraph{Majorana DM.} Self-conjugacy forbids both dipoles, since
$\bar{\chi}\sigma^{\mu\nu}\chi \equiv 0$, so the lowest-dimension operators
permitted are the axial anapole (dimension-6) and the fermionic Rayleigh
operators (dimension-7):
\begin{align}
    \mathcal{O}_{\rm AP}^{M}
    &\;=\;
    \frac{c_a}{\Lambda^{2}}\,\bar{\chi}\gamma^{\mu}\gamma^{5}\chi\,\partial^{\nu}F_{\mu\nu},
    \label{eq:O_anapole_majorana}\\[2pt]
    \mathcal{O}_{\rm R}^{M}
    &\;=\;
    \frac{c_s}{4\Lambda^{3}}\,\bar{\chi}\chi\,F_{\mu\nu}F^{\mu\nu}
    \;+\;
    \frac{c_p}{4\Lambda^{3}}\,\bar{\chi}i\gamma^{5}\chi\,F_{\mu\nu}\tilde{F}^{\mu\nu},
    \label{eq:O_rayleigh_majorana}
\end{align}
following the normalisation of Weiner \& Yavin~\cite{WeinerYavin2012RayDM},
whose factor of $1/4$ we retain so that our $\Lambda$ coincides with the
Rayleigh scale $\Lambda_R$ of that work and of the gamma-ray-line literature
built on it. That identification fixes a convention worth stating, since the
Majorana and Dirac cases are otherwise easy to mismatch. For a self-conjugate
field the single insertion of $\bar\chi\Gamma\chi$ between two external $\chi$
legs admits two contractions rather than one, but the resulting factor is common
to elastic scattering and to the annihilation process that defines $\Lambda_R$,
so it cancels in the elastic cross section at fixed $\Lambda_R$. We therefore use
one squared amplitude for both cases at equal Wilson coefficient, which is the
convention in which $\Lambda = \Lambda_R$ holds. Adopting instead a
literal-Lagrangian convention with an explicit Majorana $1/2$ would shift
$\Lambda$ by $4^{1/6} = 0.10$~dex and break that identification, and is a
relabelling rather than a change of physics. The Rayleigh operators are generated by a charged loop coupling to
two photons, in direct analogy with the Higgs-to-diphoton decay
$H\to\gamma\gamma$~\cite{Marciano2012Hgg}.

The vector current $\bar{\chi}\gamma^{\mu}\chi$ also vanishes identically for a
self-conjugate field, so no independent Majorana charge-radius operator exists.
The dimension-6 fermion current is the axial anapole above, sometimes called
the axial charge radius in the DM-EFT literature. Both names refer to
$\mathcal{O}_{\rm AP}^{M}$ and are used interchangeably here. The $CP$-violating cross-structures $\bar{\chi}\chi F_{\mu\nu}\tilde{F}^{\mu\nu}$
and $\bar{\chi}i\gamma^{5}\chi F_{\mu\nu}F^{\mu\nu}$ are also permitted, but are
dropped throughout as each is the photon-dual of a retained operator and carries
the same polarisation-summed matrix element. Replacing $F_{\mu\nu}F^{\mu\nu}$ by
$F_{\mu\nu}\tilde{F}^{\mu\nu}$ leaves the unpolarised two-photon rate unchanged,
so the scan of Eq.~\eqref{eq:O_rayleigh_majorana} already covers them. The two
sets differ only in the linear-polarisation correlations of the scattered
photons, the observable discussed in Sec.~\ref{sec:conclusions}.

\paragraph{Dirac DM.}  For a Dirac fermion, both charge conjugation, $C$, and its combination with parity, $CP$, permit a richer basis. In addition to the Majorana operators above, two dimension-5 dipole operators are now allowed,
\begin{align}
    \mathcal{O}_{\rm MD}^{D}
    &\;=\;
    \frac{c_M}{\Lambda}\,\bar{\psi}\sigma^{\mu\nu}\psi\,F_{\mu\nu},
    \label{eq:O_magneticdipole_dirac}\\[2pt]
    \mathcal{O}_{\rm ED}^{D}
    &\;=\;
    \frac{c_E}{\Lambda}\,\bar{\psi}i\sigma^{\mu\nu}\gamma^{5}\psi\,F_{\mu\nu},
    \label{eq:O_electricdipole_dirac}
\end{align}
the magnetic and electric dipole moments, respectively. These are forbidden for
self-conjugate Majorana fermions because
$\bar{\chi}\sigma^{\mu\nu}\chi \equiv 0$. At dimension 6, the Dirac case
additionally admits the charge-radius operator,
\begin{equation}
    \mathcal{O}_{\rm CR}^{D}
    \;=\;
    \frac{c_r}{\Lambda^{2}}\,\bar{\psi}\gamma^{\mu}\psi\,\partial^{\nu}F_{\mu\nu},
    \label{eq:O_chargeradius_dirac}
\end{equation}
the vector counterpart of the axial anapole of
Eq.~\eqref{eq:O_anapole_majorana}.

\paragraph{Which operators scatter real photons.} Not every operator in this
catalogue contributes to elastic $\gamma\chi\to\gamma\chi$ at tree level. The
anapole and charge-radius structures couple through $\partial^{\nu}F_{\mu\nu}$,
whose momentum-space vertex is the transverse projector
$-(k^{2}\varepsilon_{\mu} - k_{\mu}\,k\!\cdot\!\varepsilon)$. For an on-shell
external photon ($k^{2}=0$, $k\!\cdot\!\varepsilon=0$) this vanishes
identically, so the tree-level elastic amplitude of $\mathcal{O}_{\rm AP}$, and
of the Dirac charge radius, $\mathcal{O}_{\rm CR}^{D}$, is exactly zero. These are $k^{2}$-dependent form
factors that couple DM to virtual photons, that is, to charged matter, rather
than to free radiation, and they remain constrained by direct detection and by
colliders. Their two-real-photon coupling first arises at loop level through the
DM polarisability~\cite{Latimer2017AnapolePhotons,KavanaghPanciZiegler2019FaintLight},
so photon attenuation is blind to them at tree level at any coefficient and any
exposure. The operators with non-vanishing tree-level real-photon amplitudes are
therefore the two Dirac dipoles and the Rayleigh family.

\paragraph{Amplitudes.} The dipole case carries one further structural feature. Because the dipole vertex couples a single photon to the DM line, elastic $\gamma\chi\to\gamma\chi$ requires two operator insertions ($s$- and $u$-channel $\chi$ exchange, the Compton topology), so the cross section scales as $\sigma \propto c^{4}/\Lambda^{4}$ rather than the single-insertion $c^{2}/\Lambda^{2}$ suggested by naive dimension counting, and carries no factor of $\alpha_{\rm em}$. 
The exact spin-averaged squared amplitude is
\begin{equation}
    \overline{|\mathcal{M}|^{2}}
    \;=\;
    4\mu^{4}\!\left[\,-ab \;-\; 2m_\chi^{2}\,t \;+\; \frac{2m_\chi^{4}\,t^{2}}{ab}\,\right],
    \label{eq:dipole_compton_m2}
\end{equation}
with $\mu = 2c_{M,E}/\Lambda$, $a = s - m_\chi^{2}$, $b = u - m_\chi^{2}$, reducing at low energy to $\sigma \to 4\mu^{4}E_\gamma^{2}/3\pi$, the magnetic analogue of Thomson scattering with the parametric $\sigma \sim \mu^{4}E_\gamma^{2}$ scaling of Ref.~\cite{Sigurdson2004DipoleDM}. 
The magnetic and electric dipoles yield identical spin-averaged real-photon amplitudes (electromagnetic duality rotates one into the other), so photon attenuation cannot distinguish them.

The two $CP$-conserving fermionic Rayleigh operators, written for a neutral
fermion, $\Psi$, that may be either Majorana or Dirac, and the scalar Rayleigh
operator give
\begin{align}
    \bar{\Psi}\Psi\,F_{\mu\nu}F^{\mu\nu}:\quad
    & \overline{|\mathcal{M}|^{2}} = \frac{c_s^{2}}{4\Lambda^{6}}\,(4m_\chi^{2}-t)\,t^{2}, \\[2pt]
    \bar{\Psi}i\gamma^{5}\Psi\,F_{\mu\nu}\tilde{F}^{\mu\nu}:\quad
    & \overline{|\mathcal{M}|^{2}} = \frac{c_p^{2}}{4\Lambda^{6}}\,(-t)^{3}, \\[2pt]
    \phi^{2}\,F_{\mu\nu}F^{\mu\nu}:\quad
    & \overline{|\mathcal{M}|^{2}} = \frac{16c_\phi^{2}}{\Lambda^{4}}\,t^{2},
    \label{eq:rayleigh_amps}
\end{align}
with $t$ the Mandelstam momentum transfer evaluated in the DM rest frame,
Eq.~\eqref{eq:mandelstam}, and $m_\chi$ the mass of whichever candidate is under
discussion. When both coefficients are non-zero the two fermionic amplitudes add
incoherently, the interference term vanishing in the polarisation-summed
amplitude, and we quote that sum as the combined Rayleigh case. The differential
cross section in the DM rest frame is
$\dd\sigma/\dd\Omega = \overline{|\mathcal{M}|^{2}}\,(\omega\p/\omega)^{2}/(64\pi^{2}m_\chi^{2})$,
with $\omega\p/\omega$ the Compton recoil ratio of
Eq.~\eqref{eq:compton_shift}. Integrating over solid angle gives
$\sigma_{\rm tot}(E_\gamma, m_\chi)$.

Wilson coefficients are fixed at the benchmark values $c_s = c_p = 1$ (Rayleigh operators), $c_M = c_E = 1$ (dipoles), and $c_\phi = 1$ (scalar Rayleigh). 
Rescaling any coefficient $c \to \kappa c$ translates the corresponding sensitivity contour rigidly along the $\Lambda$ axis. 
For the dipoles, the invariant combination is $c/\Lambda$, so the contour shifts by the full factor $\kappa$.

For each operator with a non-vanishing real-photon amplitude we compute the
differential cross section $\dd\sigma/\dd\Omega$ for elastic
$\gamma\chi\to\gamma\chi$ at tree level in the EFT, average over initial DM
spins and photon polarisations, integrate over the scattering angle to obtain
$\sigma_{\rm tot}(E_\gamma, m_\chi)$, and assemble the optical depth through
Eq.~\eqref{eq:tau_general_revisit}.

The geometry is set by the line of sight. The source is the measured halo
emission, whose intensity follows the $\rho^{2}$ template and fills the
region of interest; the scatterer is the DM along each sightline, entering
only through the column $J \propto \int\rho_\chi\,\dd l$. The two occupy the
same region but are distinct, and need not be the same particle: in the
two-component picture of Sec.~\ref{sec:uv_translations} a heavy component
sources the emission through annihilation while a light component carries the
scattering, a freedom parametrised by $f_{\rm scat}$
(Sec.~\ref{subsec:halo_observable}). We measure the scattering angle $\theta$
from the photon's original direction of travel toward the detector, so an
unscattered photon has $\theta = 0$, and a single scatter displaces the
photon's apparent sky position by $\theta$. If the displaced position still
lies inside the ROI, the photon remains in the measured flux;
if not, it is lost. The recovery weight $w(\theta)$ is the probability of the
first outcome, averaged over emission points weighted by the $\rho^{2}$
template and over the azimuth of the deflection; it falls from unity at
$\theta = 0$ to one half at $\theta \simeq 48\degree$, the fall at small
angles set by the sharp $|b| = 10\degree$ disk edge of the box ROI
(Sec.~\ref{subsec:halo_dataset}).

What counts as a scatter differs between the two observables, and the angular
integral is taken accordingly. For the reshaping observable it runs over the
full range: a photon deflected into the near-forward direction loses a
negligible fraction of its energy and is returned to its own energy bin by the
redistribution kernel of Eq.~\eqref{eq:K_app}, so near-forward scattering
cancels between the survival and in-scatter terms of
Eq.~\eqref{eq:atten+reshape} by construction, and excluding it from
$\sigma_{\rm tot}$ as well would remove it from the bookkeeping twice.

The pure-attenuation observable keeps only photons that never scattered, so it
needs a criterion for when a scattered photon counts as removed from the
measured spectrum.
A photon leaves the measured spectrum if its energy shifts by more than the
width of its bin, or if it is deflected out of the region of interest. Writing
$x \equiv 1-\cos\theta$ and $u \equiv (\omega/m_\chi)\,x$, the Compton relation
of Eq.~\eqref{eq:compton_shift} gives a fractional energy loss
$\Delta\omega/\omega = u/(1+u)$, so a loss exceeding a fraction $f$ of the
photon energy corresponds to
\begin{equation}
    u_{E} = \frac{f}{1-f}, \qquad
    x_{E}(\omega, m_\chi) = \frac{m_\chi}{\omega}\,\frac{f}{1-f}.
    \label{eq:removal_criterion}
\end{equation}
We set $f = 0.231$, the fractional loss that carries a photon from a bin centre
to the lower edge of its bin on the native logarithmic grid of geometric ratio
$1.69$ (Sec.~\ref{subsec:halo_observable}). Angular escape, by contrast, is not
a step: it is weighted by the recovery fraction $w(\theta)$ defined above,
with half-recovery at $\theta \simeq 48\degree$.

The two channels act differently and are therefore combined rather than chosen
between. Energy migration is a genuine step at the bin edge and is applied as
one, while angular escape is integrated against its weight,
\begin{equation}
    \sigma_{{\rm rem},i} = \bigl[F_i(u_\star) - F_i(u_{\max})\bigr]
    - \int_{0}^{u_\star} \frac{{\rm d}F_i}{{\rm d}u}\,
      \bigl[1-w(\theta(u))\bigr]\,{\rm d}u ,
    \label{eq:removal_two_channel}
\end{equation}
with $u_\star \equiv \min(u_{E},\,u_{\max})$ and
$\cos\theta(u) = 1 - (m_\chi/\omega)\,u$, so that nothing is integrated beyond a
full backscatter. The second term is cut off exactly where the first begins, so
no photon is counted twice, and both vanish once $u_{E} \geq u_{\max}$: above
$m_\chi \simeq 6.7\,\omega$ even a full backscatter shifts a photon by less than
its bin width and the removal channel is closed.

Replacing the weight by a hard cut, $w \to \Theta(\theta_{\rm ROI}-\theta)$,
collapses Eq.~\eqref{eq:removal_two_channel} to
$F_i(\min(u_{E},\,u_{\rm ROI})) - F_i(u_{\max})$, the single-cut form of
Eq.~\eqref{eq:sigma_from_F} with $x_{\min} = \min(x_{E},\,x_{\rm ROI})$,
$x_{\rm ROI} = 1-\cos\theta_{\rm ROI}$ and
$u_{\rm ROI} = (\omega/m_\chi)\,x_{\rm ROI}$. In that form the two criteria
cross at $m_\chi = \omega\,x_{\rm ROI}(1-f)/f$, about $190~\mathrm{GeV}$, below
which energy migration binds and the result is independent of
$\theta_{\rm ROI}$; taking $\theta_{\rm ROI}$ at half or twice $48\degree$
shifts the contours by less than $0.02$~dex. Unlike a fixed angular cut,
$x_{E}$ scales with $m_\chi/\omega$, so the energy criterion follows the Compton
kinematics rather than being imposed on top of them.

\paragraph{Cross sections.} Writing $D \equiv m_\chi + \omega x$, the angular
integral can be performed in closed form for every surviving operator. Each total
cross section is the difference of a single function evaluated at the two
endpoints,
\begin{equation}
\begin{split}
    \sigma_{{\rm tot},i}(\omega, m_\chi) \;=\; F_i(u_{\min}) \,-\, F_i(u_{\max}), \\
    u \equiv \frac{\omega}{m_\chi}x, \qquad u_{\max} = \frac{2\omega}{m_\chi},
\end{split}
\label{eq:sigma_from_F}
\end{equation}
with $u_{\min} = (\omega/m_\chi)x_{\min}$, where $x_{\min}=0$ for the reshaping
observable, while for the attenuation observable the two-channel removal
weighting of Eq.~\eqref{eq:removal_two_channel} replaces the single endpoint and
reduces to $x_{\min} = \min(x_{E},\,x_{\rm ROI})$ in the hard-cut limit. The
functions are
\begin{align}
    F_{\rm scal}(u) &= \frac{2c_\phi^{2}m_\chi\omega\,\bigl(1+3u+3u^{2}\bigr)}
                             {3\pi\Lambda^{4}(1+u)^{3}},
    \label{eq:F_scalar}\\[4pt]
    F_{\rm odd}(u)  &= \frac{c_p^{2}m_\chi^{2}\omega^{2}(1+2u)\bigl(1+2u+2u^{2}\bigr)}
                             {64\pi\Lambda^{6}(1+u)^{4}},
    \label{eq:F_odd}\\[4pt]
    F_{\rm even}(u) &= \frac{c_s^{2}m_\chi^{3}\omega\,P(u)}
                             {192\pi\Lambda^{6}(1+u)^{4}},
    \label{eq:F_even}\\[4pt]
    F_{\rm dip}(u)  &= \frac{m_\chi^{2}\mu^{4}}{8\pi r}
      \biggl[\frac{(3+4u)+2r(1+2u)+2r^{2}}{(1+u)^{2}} \nonumber\\[2pt]
    &\qquad\qquad\qquad\quad + \; 2\ln(1+u)\biggr],
    \label{eq:F_dipole}
\end{align}
where $r \equiv \omega/m_\chi$, and
\begin{equation}
\begin{split}
    P(u) = \;&8\bigl(1+4u+6u^{2}+3u^{3}\bigr) \\
         &+ 3r\bigl(1+4u+6u^{2}+4u^{3}\bigr).
\end{split}
\label{eq:F_even_poly}
\end{equation}
The combined Rayleigh cross section is $F_{\rm even}+F_{\rm odd}$ evaluated
between the same limits. In the regime $\omega \ll m_\chi$, these reduce to
$\sigma \propto \omega^{2}$ for the dipoles, $\omega^{4}$ for the scalar and for
$\bar{\Psi}\Psi F_{\mu\nu}F^{\mu\nu}$, and $\omega^{6}$ for
$\bar{\Psi}i\gamma^{5}\Psi F_{\mu\nu}\tilde{F}^{\mu\nu}$, the scalings of
Sec.~\ref{subsec:eft_results}. Every $F_i$ is finite at $x=0$, so the reshaping
cross section is well defined without a cut, and the dipole form reproduces the
low-energy Thomson-like limit $\sigma \to 4\mu^{4}\omega^{2}/3\pi$ there.

\paragraph{Energy and cutoff scalings.} Every surviving operator produces a
total cross section that scales as a fixed power of the photon energy in the
heavy-scatterer regime $\omega \ll m_\chi$, where the Compton recoil is
negligible and the momentum transfer reduces to
$|t| \simeq 2\omega^{2}(1-\cos\theta)$, independent of $m_\chi$. Each power of
$|t|$ in the spin-averaged squared amplitude contributes two powers of
$\omega$, and the phase-space measure and the flux factor of
Eq.~\eqref{eq:dsigma_dt} supply compensating factors of $\omega^{2}$, so the
exponent can be read from the squared amplitude alone. Writing
$\sigma_{{\rm tot},i} \propto \omega^{\,k_i}$, we find $k_i = 2$ for the
dimension-5 dipoles, $k_i = 4$ for the dimension-6 scalar Rayleigh operator
and the parity-even dimension-7 Rayleigh operator, and $k_i = 6$ for the
parity-odd dimension-7 operator. The two dimension-7 operators of
Eq.~\eqref{eq:O_rayleigh_majorana} carry the same field strengths, and hence
the same two powers of $t$, meaning they differ only through the non-relativistic
reduction of the DM bilinear. The scalar bilinear is unsuppressed,
$\bar\chi\chi \to 2m_\chi$, so the parity-even squared amplitude retains the
term $\propto m_\chi^{2}t^{2}$ of Eq.~\eqref{eq:rayleigh_amps}; the
pseudoscalar bilinear is of order $\sqrt{|t|}$, so the parity-odd squared
amplitude begins at $(-t)^{3}$. That single extra power of $|t|$ is the
difference between $k_i = 4$ and $k_i = 6$.

The dependence on $\Lambda$ is set instead by the number of operator
insertions. The Rayleigh operators contribute at a single insertion, giving
$\sigma_{{\rm tot},i} \propto \Lambda^{-2(d_i-4)}$. A dipole cannot mediate
elastic $\gamma\chi$ scattering at a single insertion, since the operator is
linear in $F_{\mu\nu}$ and carries only one photon leg. The process therefore proceeds
through the two-insertion $s$- and $u$-channel Compton diagrams, so
$\sigma_{\rm tot} \propto \Lambda^{-4}$, even though $d_i = 5$. That is the same power of $\Lambda$ as the dimension-6 scalar Rayleigh operator. Operator dimension therefore
fixes neither exponent on its own, and the surviving families are separated by
their energy dependence rather than by their cutoff scaling. That
operator-specific rise with $\omega$ makes high-energy gamma-ray data, rather
than low-energy photon probes, the natural probe of this class of interaction.
It is also what makes the bounds of Sec.~\ref{sec:halo_limits}
operator-resolved rather than a single limit on a generic cross section.

%------------------------------------------------------------------
\subsection{Validity and Observability Criteria}
\label{subsec:eft_criteria}
%------------------------------------------------------------------
\begin{figure*}[htbp]
    \includegraphics[width=\textwidth]{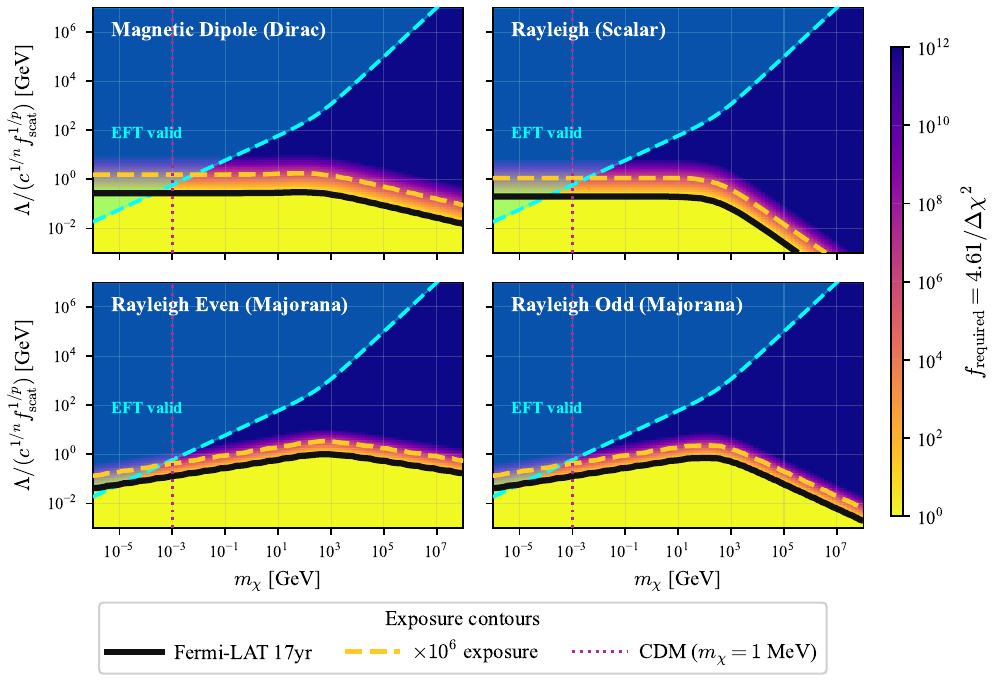}
\caption{
Exposure-multiplier sensitivity maps
    $f_{\rm required}(m_\chi, \Lambda) = 4.61/\Delta\chi^{2}_{\rm current}$
    across the $(\Lambda, m_\chi)$ parameter space for the four distinct
    surviving cases, in panel order: the Dirac dipole (the electric dipole gives an
    identical spin-averaged real-photon amplitude and is not shown separately),
    the scalar Rayleigh, and the two fermionic Rayleigh operators of
    Eq.~\eqref{eq:O_rayleigh_majorana}, which apply to Dirac and Majorana DM
    alike. The vertical axis
    is the rescaled cutoff $\Lambda/(c^{1/n}f_{\rm scat}^{1/p})$, with
    $n = d-4$ and $p$ the power of $\Lambda$ in the cross section ($p = 4$ for
    the dipole and scalar Rayleigh, $6$ for the dimension-7 operators). The
    first factor makes a common rescaling of the Wilson coefficients shift every
    panel identically; the second absorbs the fraction $f_{\rm scat}$ of the
    halo column carried by the scattering species, exactly degenerate with it
    (Sec.~\ref{subsec:halo_observable}). Panels are drawn at the benchmark
    $c = f_{\rm scat} = 1$. The
    fixed factor of $1/4$ carried by the fermionic Rayleigh operators and not by
    the scalar (Sec.~\ref{subsec:eft_catalogue}) is not absorbed by this choice,
    and displaces the scalar panel relative to the other three by $0.30$~dex.
    Each panel gives the multiplicative exposure boost needed to reach
    $90\%$~CL detectability against the \textit{Fermi}--LAT pixel-level
    Galactic-centre halo posterior of Sec.~\ref{sec:halo_limits}. Overlaid
    contours are drawn at $f_{\rm required} = 1$ (solid black, the current
    17\,yr baseline) and $f_{\rm required} = 10^{6}$. The cyan-shaded wedge is the EFT-valid region bounded by
    $\Lambda^{2} \geq \max(s_{\max}, |t|_{\max})$. The vertical dotted magenta line
    marks the CDM mass floor $m_\chi = 1~\mathrm{MeV}$.}
\label{fig:eft_grid}
\end{figure*}

For a given operator and a given point in the $(\Lambda, m_\chi)$ plane, two
criteria must be satisfied simultaneously for the predicted signal to be both
meaningful and observable.

\paragraph{EFT validity.} The effective description is reliable only where every
kinematic invariant of the scattering process lies below the cutoff,
\begin{equation}
    \Lambda^{2} \;\geq\; \max\bigl(s_{\max},\,|t|_{\max}\bigr),
    \label{eq:eft_validity}
\end{equation}
with, in the DM rest frame,
\begin{equation}
\begin{split}
    s_{\max}   &= m_\chi^{2} + 2 m_\chi \omega_{\max}, \\
    |t|_{\max} &= \frac{4\,\omega_{\max}^{2}}{1 + 2\omega_{\max}/m_\chi},
\end{split}
\label{eq:s_t_max}
\end{equation}
both evaluated at $\omega_{\max}$, the highest photon energy retained in the
fit. Since $|t|_{\max} = (s-m_\chi^{2})^{2}/s < s$ for all elastic kinematics,
the binding condition is always $s_{\max}$: for $m_\chi \gg \omega_{\max}$ it
reduces to $\Lambda \gtrsim m_\chi$, and for $m_\chi \ll \omega_{\max}$ to
$\Lambda \gtrsim \sqrt{2 m_\chi \omega_{\max}}$. Identifying $\Lambda$ with the
mediator mass through $s$ is the completion-agnostic worst case, and it is
enforced by perturbative unitarity: a contact operator yields partial-wave
amplitudes growing with $s$, violating unitarity at
$\sqrt{s} \sim \sqrt{4\pi}\,\Lambda$ whatever the
completion~\cite{KavanaghPanciZiegler2019FaintLight}, and
$\sqrt{s} \geq m_\chi$ always; the wedge adopts the stricter unit-coefficient
form of that condition. Which invariant controls validity for a specified
completion depends on the channel in which the mediator
appears~\cite{Busoni2014EFTValidity}: for a $t$-channel completion, where the
momentum flowing through the mediator is $\sqrt{|t|}$, the weaker $|t|$
condition applies, as it does for the completions of
Sec.~\ref{sec:uv_translations}, and no conclusion of this work changes under
that relaxation, so the wedge as drawn is conservative. Adopting the unitarity
normalisation instead, $\Lambda^{2} \geq \max(s_{\max},|t|_{\max})/4\pi$,
would lower the wedge by $\sqrt{4\pi} \simeq 3.5$ ($0.55$~dex) and raise the
wedge-entry masses of Sec.~\ref{subsec:eft_results} by up to a factor of
$4\pi$, to roughly $3.0~\mathrm{MeV}$ for the dipoles, $1.6~\mathrm{MeV}$ for
the scalar Rayleigh and $0.15$--$0.24~\mathrm{MeV}$ for the dimension-7
operators. Every shifted crossing remains inside the thermally excluded band
of Sec.~\ref{subsec:halo_results}: the dimension-7 values sit below the big-bang nucleosynthesis (BBN)
floor, the dipole and scalar values within the $N_{\rm eff}$ argument, which
applies to any species annihilating after neutrino decoupling, and the dipole
exclusion admits no perturbative completion under either convention. The
choice therefore moves the wedge but not the conclusions, and we keep the
stricter unit-coefficient form in the figures.

 Because the surviving operators have steeply rising cross sections,
$\sigma \propto \omega^{k}$ with $k = 2$--$6$, the test statistic is necessarily
dominated by the highest retained bin, and $\omega_{\max}$ is therefore the
controlling analysis choice rather than an incidental one. It enters twice and
in opposite directions: it sets the validity wedge, which scales as
$\sqrt{\omega_{\max}}$, and it sets the reach, since discarding the top of the
band removes the bins that carry most of the signal. The three highest
halo-posterior bins carry fractional uncertainties of $35\%$, $105\%$ and
$180\%$, the upper two consistent with zero flux, so retaining them out to
$814~\mathrm{GeV}$ imposes the most restrictive validity requirement the dataset
can support. Accordingly, we choose to truncate the spectrum at $200~\mathrm{GeV}$, giving
$\omega_{\max} = 168.9~\mathrm{GeV}$, the centre of the highest surviving bin.
We additionally drop the lowest positive bin, at $2.55~\mathrm{GeV}$: it carries
a $21\%$ fractional uncertainty and, because the surviving operators scale as
$\sigma \propto \omega^{k}$ with $k \geq 2$, it contributes negligibly to the
statistic while being the bin most exposed to source-model mismodelling at the
low-energy end. The two cuts together retain 8 of the 12 positive bins and move
the inverse-variance-combined statistical uncertainty from $2.55\%$ to
$2.58\%$.

That truncation is not free, and its cost is larger in the statistic than in the
quoted limit. It removes $65$--$71\%$ of the raw $\Delta\chi^{2}$ for
the dimension-7 operators, which compresses to a $16\%$ loss in reach, more
than the naive $\Lambda^{1/12}$ scaling would give because profiling the source
normalisation makes the statistic sub-quadratic in $\tau$ ($5.2\%$ for the dipoles and $5.7\%$ for the scalar
Rayleigh, the two being comparable). These figures are quoted from refined
$\Lambda$ grids. The production grid samples $\Lambda$ at $26\%$ per step, which
is coarser than the displacements themselves, so each was re-measured at
successively finer resolution until the value stabilised and the displacement
spanned several grid cells. The three dimension-7 operators agree to better than
$0.2$ percentage points once resolved; the apparent spread between them at
production resolution was grid quantisation. Against that, the low-mass wedge moves down by
$\sqrt{814/168.9} = 2.20$, or $0.34$~dex, for
$m_\chi \lesssim 1~\mathrm{GeV}$. The gain falls to $0.30$~dex at
$m_\chi = 100~\mathrm{GeV}$ and $0.15$~dex at $1~\mathrm{TeV}$, vanishing once
the criterion becomes mass-dominated. This is the cyan dashed curve in
Fig.~\ref{fig:eft_grid}; the cyan-shaded region above it is EFT-valid, and
points below it are EFT-invalid.
 
Two conservative choices are implicit here. First, validity is imposed at the
single highest retained bin for the entire dataset. The standard self-consistent
alternative is a per-bin truncation retaining only bins with
$\sqrt{s} < \Lambda$ in the test statistic. That alternative requires no
separate scan: truncation removes only non-negative terms from
Eq.~\eqref{eq:delta_chi2_general}, so
$\Delta\chi^{2}_{\rm trunc}(m_\chi,\Lambda) \leq \Delta\chi^{2}(m_\chi,\Lambda)$
pointwise, and since the untruncated contour lies below the validity wedge
throughout the cold dark matter (CDM) range the per-bin-truncated scan excludes nothing there
by construction. Second, the criterion identifies $\Lambda$ with the mediator
mass, which holds at the benchmark $c = 1$. Since the data constrain only the
combination $c^{1/n}/\Lambda$, a strongly coupled completion ($c \to 4\pi$)
shifts the wedge comparison upward by $(4\pi)^{1/n}$, which is $1.1$~dex for the
dimension-5 dipoles, and $0.55$ ($0.36$)~dex at $n = 2$ ($3$). The fixed
normalisation difference between the scalar and fermionic Rayleigh conventions
acts the same way at smaller amplitude, $0.30$ and $0.20$~dex respectively, and
is comparable to the wedge margins it displaces. The $c = 1$ benchmark is
nevertheless the appropriate one: a dipole generated by a charged loop carries
$c \sim e g^{2}/16\pi^{2} \ll 1$, and the loop-induced Rayleigh operators
likewise have $c \sim \alpha/\pi \ll 1$, for which the natural exclusion is
correspondingly weaker than the $c = 1$ contours shown, not stronger.
 
%% [REMOVED 2026-08-06] Partial-wave unitarity guide. All four legacy expressions
%% were dimensionally inconsistent (returning Lambda in GeV^1/2 or GeV^1/3), and a
%% correctly derived guide, Lambda_unit = sqrt(s)*(C/8pi)^(1/n) against
%% Lambda_kin = sqrt(kappa*s), is a constant factor BELOW the kinematic curve at
%% c = 1 and therefore never binds. Curve also removed from Fig. 2.
%% Optional footnote: at the benchmark coupling the partial-wave bound is weaker
%% than the kinematic criterion by a constant factor, becoming relevant only for
%% C >~ 8pi.

\paragraph{Observability.} Rather than adopt a single $\tau$ threshold, we
evaluate the test statistic of Eq.~\eqref{eq:delta_chi2_general} for each
operator across the $(\Lambda, m_\chi)$ plane, taking
$\Phi_{\rm src} = \Phi_{\rm data}$, the \textit{Fermi}--LAT pixel-level
Galactic-centre halo posterior of Sec.~\ref{subsec:halo_dataset}, and $\sigma_i$
as the corresponding bin-level statistical uncertainties. The exposure
multiplier $f_{\rm required}(m_\chi, \Lambda)$ of Eq.~\eqref{eq:f_required} then
has a direct reading. The $f_{\rm required} = 1$ contour separates the region in
which the present exposure already suffices to resolve the distortion from the
region requiring more. The second contour, $f_{\rm required} = 10^{6}$, is drawn
not as a projection for any planned instrument but to show how little the
boundary moves under an exposure increase far beyond anything foreseeable.

That reading holds only while the measurement is statistics-limited, and we do
not attempt to mark where it ceases to be so. The dominant systematic, template composition in the diffuse
model~\cite{StenhouseGhagDeppisch2026Totani}, does not integrate down, so a
ceiling on $f_{\rm required}$ exists in principle; locating it requires the
energy-dependent component of that systematic, since a fully correlated
normalisation error is absorbed by the profiled $A$ and does not degrade a
shape measurement. That component is not separable from the inputs used here.
Values of $f_{\rm required}$ well above unity should therefore be read as
requirements rather than as schedules. Translating a multiplier into a specific instrument
requires more than an exposure ratio, since effective area, energy band, angular
resolution and background systematics all differ; Sec.~\ref{subsec:halo_results}
sets out how the wedge and the contour respond to a change of band, which is the
comparison that determines whether a given instrument helps.
\paragraph{Cosmological reference.}  We do not impose $m_\chi \geq 1~\mathrm{MeV}$ as a hard cut on the displayed range, making the EFT analysis agnostic about the cosmological history of $\chi$. Signals below this reference (dotted magenta in Fig.~\ref{fig:eft_grid}) should be read as relevant to warm, light, or otherwise non-CDM scenarios rather than as physically excluded by the CDM prior alone;
Sec.~\ref{subsec:halo_results} shows this window is nonetheless closed by
thermalisation and BBN for any standard thermal history.

%------------------------------------------------------------------
\subsection{Results}
\label{subsec:eft_results}
%------------------------------------------------------------------

Fig.~\ref{fig:eft_grid} shows $f_{\rm required}(m_\chi, \Lambda)$ across the
full operator basis. In each panel the $f_{\rm required} = 1$ contour is the
current $90\%$~CL \textit{Fermi}--LAT exclusion, and the colour scale gives the
exposure multiplier needed to reach any other point in the plane.
The further contour, at $f_{\rm required} = 10^{6}$, marks where the
exclusion would land under an exposure increase far beyond anything
foreseeable.

Among the surviving operators the cross section is suppressed by $\Lambda^{-4}$
for the dipoles, which enter through two insertions of the dimension-5 operator,
and for the dimension-6 scalar Rayleigh operator, and by $\Lambda^{-6}$ for the
dimension-7 Rayleigh family. At the current \textit{Fermi}--LAT exposure
($f_{\rm required}=1$, solid black) the dipole panels exclude up to
$\Lambda \simeq 0.32~\mathrm{GeV}$, peaking at $m_\chi \simeq 60~\mathrm{GeV}$.
The scalar Rayleigh reaches $\Lambda \simeq 0.21~\mathrm{GeV}$ at
$m_\chi \simeq 0.22~\mathrm{GeV}$, and the dimension-7 Rayleigh panels
$\Lambda \sim 0.79$--$1.06~\mathrm{GeV}$ at
$m_\chi \sim 0.12$--$0.58~\mathrm{TeV}$.
Below $m_\chi \sim 1~\mathrm{GeV}$ the dipole contour is flat at
$\Lambda \simeq 0.29~\mathrm{GeV}$, as expected: in the $\omega \gg m_\chi$
regime $\sigma \propto \mu^{4} m_\chi \omega$, so $\tau = (J/m_\chi)\sigma$ is
independent of $m_\chi$.

These boundaries respond only weakly to added exposure. Since
$\Delta\chi^{2} \propto \sigma^{2} \propto \Lambda^{-8}$ for the dipole and
scalar-Rayleigh operators and $\Lambda^{-12}$ for the dimension-7 family, an
order of magnitude more data moves them by $10^{1/8} \approx 1.3$ and
$10^{1/12} \approx 1.2$, or $0.13$ and $0.08$ dex in $\Lambda$. This is the
reason for the second contour of Fig.~\ref{fig:eft_grid}: even a boost of
$10^{6}$ in exposure moves the dimension-7 boundary by only
$10^{6/12} \approx 3$, from $\Lambda \sim 0.79$--$1.06~\mathrm{GeV}$ to
$\Lambda \approx 3.4~\mathrm{GeV}$. The weak dependence is a property of the
observable rather than of this dataset, and no realistic exposure changes it.

No operator in the scan produces a current-exposure testable region above the
CDM reference line, $m_\chi = 1~\mathrm{MeV}$, within the EFT-valid wedge. At
that reference the contours clear the wedge by between $0.30$~dex for the dipoles
and $0.64$~dex for the parity-even and parity-odd Rayleigh operators.
Every contour does cross into the wedge at lower mass, the dipoles below
$m_\chi \simeq 240~\mathrm{keV}$, the scalar Rayleigh below $130~\mathrm{keV}$
and the dimension-7 Rayleigh operators below $12$--$19~\mathrm{keV}$. These
crossings are where two shallow curves meet and are correspondingly sensitive to
the scan resolution, so we quote them to two significant figures. Refining the $\Lambda$ grid
twelvefold moves the peak $\Lambda$ values downward by $1$--$5\%$, the dipoles
and the parity-odd Rayleigh operator being the most affected; the values quoted
here are from the production grid. Each lies more
than half a decade below the CDM floor, so the EFT-valid regions of the scan
pertain to warm or light DM rather than to a conventional cold relic.
Sec.~\ref{subsec:halo_results} shows that this window is nonetheless closed by
thermalisation and BBN for any standard thermal history, so it is not open
parameter space either.

The dipoles are the operators with the most natural UV origin, being generated
by a charged loop at the electroweak scale with
$c \sim e g^{2}/16\pi^{2} \ll 1$, so the $c = 1$ benchmark shown is already
generous. Magnetic and electric dipoles give identical attenuation signals, so a
detection in this channel would measure the combination $c_{M,E}/\Lambda$
without distinguishing the two moments. Separating them requires a DM--matter
channel or polarimetry.

The vanishing of the anapole amplitude yields a sharp discriminant between
Dirac and Majorana DM. For a Majorana fermion the dipole operators are
identically zero ($\bar\chi\sigma^{\mu\nu}\chi \equiv 0$) and the anapole does
not couple to free photons, so the only tree-level real-photon channel open to
Majorana DM at $d \leq 7$ is the Rayleigh family of
Eq.~\eqref{eq:O_rayleigh_majorana}, with its steep
$\sigma \propto E_\gamma^{4}$--$E_\gamma^{6}$ energy dependence. An attenuation
signal carrying the dipole's characteristic
$\tau(E_\gamma) \propto E_\gamma^{2}$ scaling would therefore exclude a single
Majorana state as the scatterer, a statement inaccessible to annihilation and
decay searches. The exclusion is of a single self-conjugate state rather than of
self-conjugacy in general: as Sec.~\ref{subsec:ew_doublet_dipole} discusses, a
pseudo-Dirac pair with a small mass splitting $\delta$ carries a transition
dipole and reproduces the Dirac scaling for $E_\gamma \gg \delta$, so a
dipole-like signal constrains the spectrum of states rather than their
individual character. The Rayleigh channel left to Majorana DM currently reaches
only $\Lambda \lesssim 1.1~\mathrm{GeV}$.

%==================================================================
\section{Scattering Sensitivity from \textit{Fermi}--LAT Galactic Centre Data}
\label{sec:halo_limits}
%==================================================================

Sec.~\ref{sec:eft_operators} established which EFT operators could be tested
given a sufficiently dense and well-characterised photon flux. We now turn this
around. The disk-excluded halo posterior of
Ref.~\cite{StenhouseGhagDeppisch2026Totani}, resolved from 17~years of
\textit{Fermi}--LAT observations toward the Galactic centre, is the largest DM
column on the sky for which a measured spectrum exists, the cusped on-axis
sightline being removed by the disk cut. Asking how large a spectral distortion
each operator would imprint on that spectrum, relative to its measured
uncertainties, gives the region of the $(\Lambda, m_\chi)$ plane to which the
channel is already sensitive.

%------------------------------------------------------------------
\subsection{Dataset and halo posterior}
\label{subsec:halo_dataset}
%------------------------------------------------------------------

We use 17~years of \textit{Fermi}--LAT Pass~8 ULTRACLEAN
photons~\cite{Atwood2009FermiLAT,Ackermann2012FermiLikelihood} toward the
Galactic centre, analysed at the pixel level in Ref.~\cite{StenhouseGhagDeppisch2026Totani} following the
morphological template framework of Ref.~\cite{totani202520gevhalolikeexcess}.
In summary, the diffuse emission is decomposed into Galactic-diffuse,
isotropic, Loop~I, \textit{Fermi}-bubble, point-source, and spherically-symmetric
NFW-like halo templates, with the halo template normalised at the Galactic
pole. An \texttt{emcee}~\cite{emcee2013} Poisson MCMC over the template coefficients in each energy bin returns the posterior
flux, $E^{2}\dd N/\dd E$, of the halo component at $J/J_{\rm pole}=1$, denoted
$\Phi_{\rm halo}(E)$. For the morphology we consider two NFW emissivity
scalings $\rho^{p}$, a $p=2$ profile and a steeper $p=2.5$ stress test, with
the Galactic disk excluded from the ROI. This isolates the same
high-latitude, spatially and spectrally distinct halo-like component
discussed in Ref.~\cite{StenhouseGhagDeppisch2026Totani}. The corresponding pole-normalised columns are
$J_{\rm pole}^{\rho^{2}} \approx 4.0\times10^{21}~\mathrm{GeV^{2}\,cm^{-5}}$
and $J_{\rm pole}^{\rho^{2.5}} \approx 1.6\times10^{22}~\mathrm{GeV^{2.5}\,cm^{-6.5}}$.
The scattering optical depth instead uses the emissivity-weighted, ROI-averaged linear column $J = \int\rho\,\dd l$ (the $n=1$ moment, not $\rho^{2}$) of the same NFW profile. 
For the disk-excluded halo ROI ($|b|\geq10\degree$), this is $J_{\rm ROI} \approx 4.8\times10^{22}~\mathrm{GeV\,cm^{-2}}$ under the $\rho^{2}$ annihilation weighting applied throughout the scan (and $\approx 5.4\times10^{22}~\mathrm{GeV\,cm^{-2}}$ under a $\rho^{2.5}$ weighting).
This is of the same order as the constant-density Sun--Galactic-centre column of Eq.~\eqref{eq:galactic_baseline}, $J_{\rm gc} = 1.05\times10^{22}~\mathrm{GeV\,cm^{-2}}$, and well below the cusped on-axis NFW sightline that the disk cut removes.

%---
\paragraph{Why the halo (and only the halo).}
For the purpose of constraining photon--DM scattering, it suffices to treat
$\Phi_{\rm halo}(E)$ as a measured spectrum with a well-defined morphology
and statistical uncertainty, as the scattering attenuates and reshapes whatever
flux propagates along the line of sight, regardless of its origin. However,
we do not apply the scattering test uniformly across all diffuse
templates, and the reason is physical rather than pipeline-driven. The analysis
of Ref.~\cite{StenhouseGhagDeppisch2026Totani} establishes that if the 20~GeV
halo has a DM origin at all, the only interpretation that is simultaneously
consistent with (i)~the halo normalisation, (ii)~the dwarf spheroidal galaxy
$\langle\sigma v\rangle$ limits, and (iii)~the observed relic density is
low-velocity-enhanced annihilation of a $m_\chi \simeq 0.56$--$0.72~\mathrm{TeV}$
WIMP through a resonant Sommerfeld or Breit--Wigner mechanism. This
resolution is not minimal: it requires the annihilator to inhabit a
structured dark sector containing a light mediator whose mass sits between the
dwarf and Galactic velocity scales, corresponding to a sub-GeV mediator mass
for a TeV-scale annihilator~\cite{StenhouseGhagDeppisch2026Totani}.

That conclusion is not reached in isolation. Of the five independent
realisations of the 20~GeV excess published to date, the four that take the
Sommerfeld route each introduce a new state light enough to mediate a
long-range force at halo velocities: a $CP$-even scalar near $400~\mathrm{MeV}$
in a Higgs-portal dark-photon model~\cite{Yamashita2026TotaniHiggsPortal}, a
mediator below $\sim 500~\mathrm{MeV}$ in a naturally inelastic electroweak
doublet~\cite{NomuraTotani2026Higgsino}, a $p$-wave secluded sector
annihilating to $W^{+}W^{-}$~\cite{Yoshimatsu2026SecludedWW}, and a
dilaton-like scalar at $m_\varphi \simeq 12~\mathrm{GeV}$ cascading through
$\chi\bar\chi \to \varphi\varphi \to 4b$~\cite{Jho2025TotaniDilaton}. The
mediator scale these constructions favour therefore spans the sub-GeV range up
to $\mathcal{O}(10)~\mathrm{GeV}$ rather than sitting at a single value, but in
every case, a state far lighter than the annihilator is populated along the same
line of sight as the photons whose spectrum is measured.

The fifth realisation is structurally different, and for our purposes more
informative. Reference~\cite{Murayama2025BreitWigner} reaches the same halo
rate through a Breit--Wigner resonance rather than a Sommerfeld enhancement,
with the resonance sitting at $M \simeq 2m_\chi$ and no light state in the
spectrum at all. That class predicts no scatterer in the halo. A photon--DM
scattering measurement therefore separates the two families on the presence or
absence of a light component, rather than on the annihilation rate that every
one of them was constructed to reproduce.

The light mediator the Sommerfeld constructions require sits at the scale our
contours reach (Sec.~\ref{sec:eft_operators}, Fig.~\ref{fig:eft_grid}), though
reaching a scale is not the same as probing it: the dipole and Rayleigh
contours peak at sub-GeV $\Lambda$ but lie below the validity wedge there, and
a light mediator is the natural counterpart to a heavy annihilating $\chi$ at
the electroweak scale for the dimension-5 Dirac dipoles
(Sec.~\ref{subsec:ew_doublet_dipole}). The scattering test is therefore posed
on the halo template because, under the fully consistent DM interpretation of
the halo proposed in Ref.~\cite{StenhouseGhagDeppisch2026Totani}, that is where
the light mediator responsible for the scattering physically resides. Applying
the same scattering distortion to the isotropic, \textit{Fermi}-bubble, or
diffuse components would require an ad~hoc population of scatterers along lines
of sight where the DM density is not enhanced. The halo, by contrast, is the
line of sight where both the annihilator (the halo photons' source) and the
scatterer (its mediator) coexist by construction. In this sense the photon--DM
scattering search of Sec.~\ref{sec:halo_limits} is the consistency test of the
two-component dark sector picture that
Ref.~\cite{StenhouseGhagDeppisch2026Totani} arrived at from the halo DM
interpretation alone, and a discriminant between that picture and the resonant
alternative.

The astrophysical interpretation of $\Phi_{\rm halo}(E)$, whether it is
sourced by DM annihilation, decay, or an unmodelled astrophysical
foreground, is the subject of Ref.~\cite{StenhouseGhagDeppisch2026Totani}, not
of this work. What the argument above requires is only that if the
halo is taken to be sourced from DM, it is the physically distinguished region for the
scattering test. The bounds we set are independent of the interpretation and
apply to the halo template as a measured spectrum regardless.

%------------------------------------------------------------------
\subsection{Observable: attenuation and reshaping}
\label{subsec:halo_observable}
%------------------------------------------------------------------

The transport formalism of Sec.~\ref{subsec:transport} carries over
unchanged. For each operator and each point in the $(\Lambda, m_\chi)$ plane we
build the transfer matrix, the discrete form of the transport in
Eq.~\eqref{eq:atten+reshape},
\begin{equation}
    T(E\,|\,E\p) = e^{-\tau(E\p)}\,\delta(E-E\p)
    + K_{w}(E\,|\,E\p)\,\tau(E\p)\,e^{-\tau(E\p)},
    \label{eq:T_app}
\end{equation}
whose kernel is the operator-specific differential cross section of
Sec.~\ref{sec:eft_operators}, integrated against the Compton kinematics of the
kicked photon, weighted and normalised as in Eq.~\eqref{eq:K_app}. The pure-attenuation observable retains only the
survival term of Eq.~\eqref{eq:atten+reshape}, which is the appropriate treatment
when scattered photons leave the observed population rather than returning to it,
and the reshaping observable retains both terms. We test both in turn.

We evaluate $T$ on the halo posterior's native energy grid, apply it to a smooth
source spectrum, and compare $\Phi_{\rm pred}$ to the measured $\Phi_{\rm halo}(E)$ via
\begin{equation}
\chi^{2}(\Lambda, m_\chi) \;=\; \sum_{i} \frac{\bigl[\Phi_{\rm halo}(E_i) - A\,\Phi_{\rm pred}(E_i\,;\,\Lambda, m_\chi)\bigr]^{2}}{\sigma_{i}^{2}},
\label{eq:halo_chi2}
\end{equation}
profiled over the source-spectrum normalisation $A$. We exclude points at the $90\%$~CL where $\chi^{2} - \chi^{2}_{\rm min} > \Delta\chi^{2}_{90}$ with $\Delta\chi^{2}_{90} = 4.61$ for two parameters of interest. The fit runs over the pixelwise \textit{Fermi}--LAT halo posterior of Ref.~\cite{StenhouseGhagDeppisch2026Totani} on its native 13-bin logarithmic energy grid (bin centres $1.51$--$814~\mathrm{GeV}$, geometric ratio $1.69$). We retain the bins with positive posterior halo flux that survive both cuts of Sec.~\ref{subsec:eft_criteria}, eight bins with centres $4.31$, $7.28$, $12.3$, $20.8$, $35.1$, $59.2$, $100$ and $169~\mathrm{GeV}$ for the $\rho^{2}$ posterior. The lowest native bin at $1.51~\mathrm{GeV}$ carries a negative posterior median and is dropped on that ground; the $2.55~\mathrm{GeV}$ bin is dropped for the reason given there. The cost of the two cuts is quantified in Sec.~\ref{subsec:eft_criteria}; the combined statistical uncertainty of the retained bins is $2.576\%$, giving the bulk detection floor $\tau_{\rm floor} \approx 2.6\times10^{-2}$ of Sec.~\ref{subsec:datasets} ($2.81\%$ and $2.8\times10^{-2}$ for the $\rho^{2.5}$ posterior). The dominant systematic, template composition in the diffuse model, which moves the halo normalisation by $5\%$ across the 14 interstellar-emission variants of Ref.~\cite{StenhouseGhagDeppisch2026Totani}, is not added
as a separate per-bin term in this $\chi^{2}$: the only per-bin weight $\sigma_i$ is the \texttt{emcee} posterior $1\sigma$, $\sigma_i = \tfrac{1}{2}(\Phi_{\rm p84} - \Phi_{\rm p16})$, of the halo fit of Ref.~\cite{StenhouseGhagDeppisch2026Totani}, which already marginalises over the diffuse, isotropic, Loop~I, \textit{Fermi}-bubble and point-source template coefficients. The systematic's effect on the contours has instead been measured directly rather than estimated. Repeating the full scan on the disk-included posterior, a distinct fit differing from the disk-excluded one by a factor of order two in normalisation and by up to $260\%$ in a single bin, weakens the exclusion by at most $0.040$~dex, or $8.8\%$ in $\Lambda$. The two effects that make so large an input change so small an output one are both structural: the statistic is profiled over the source normalisation $A$, which absorbs a pure rescaling exactly and leaves only the shape difference, and $\Lambda$ enters the optical depth at high power, which compresses what shape difference remains. The systematic is therefore about half the band-truncation cost of the dimension-7 operators quoted in Sec.~\ref{subsec:eft_criteria}, and somewhat larger than that of the dipoles and the scalar. We quote statistical contours throughout and treat this figure as the systematic envelope.

One further assumption is implicit in the normalisation of $\tau$. We take the
scattering species to carry the whole halo column, $f_{\rm scat} = 1$. If it is
instead a subdominant component, as the two-component picture of
Sec.~\ref{sec:uv_translations} allows, then $\tau \propto f_{\rm scat}$ and the
fraction is exactly degenerate with the Wilson coefficient. The contours rescale
as $\Lambda \to \Lambda\,f_{\rm scat}^{1/4}$ for the dimension-5 and dimension-6
operators, whose cross sections scale as $\Lambda^{-4}$, and as $\Lambda \to
\Lambda\,f_{\rm scat}^{1/6}$ for the dimension-7 family. Every limit quoted here
may therefore be read at any $f_{\rm scat}$ by that rescaling, and the weak
fractional powers mean even an order of magnitude in $f_{\rm scat}$ moves the
reach by less than a factor of two.

%------------------------------------------------------------------
\subsection{Operator-by-operator limits}
\label{subsec:halo_results}
%------------------------------------------------------------------

\begin{figure*}[t]
    \centering
    \includegraphics[width=\textwidth]{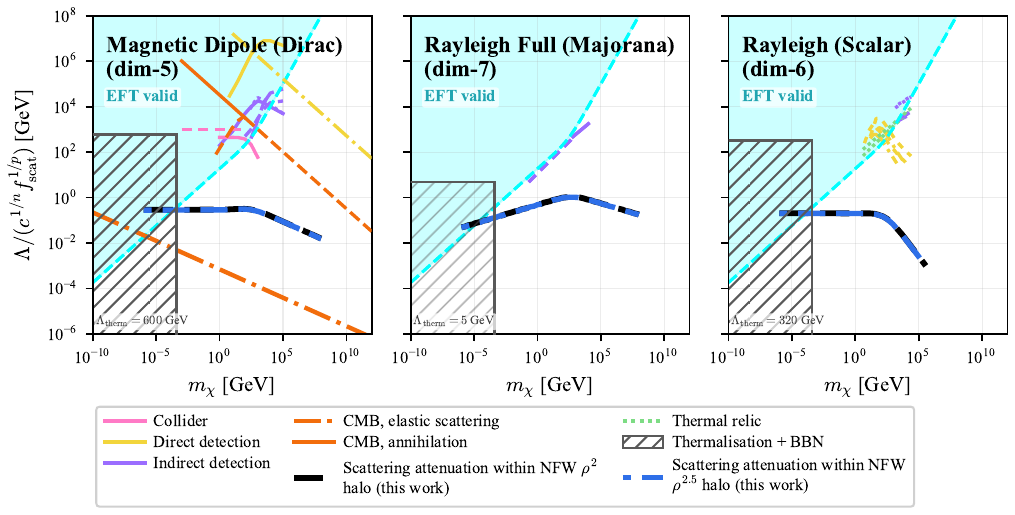}
    \caption{
        Halo-derived $90\%$~CL sensitivity contours (black and blue, NFW
        $\rho^{2}$ and $\rho^{2.5}$ profiles) for three representative operators,
        the Dirac magnetic dipole (dim-5), the Majorana Rayleigh (dim-7) and the
        scalar Rayleigh (dim-6). External constraints are digitised from the
        compilations cited in Sec.~\ref{subsec:halo_results}: the dipole panel
        carries collider, direct-detection, indirect-detection and cosmological
        curves, the scalar Rayleigh panel direct-detection, indirect-detection
        and thermal-relic curves, and the dimension-7 panel the \textit{Fermi}--LAT and H.E.S.S. line searches only. The thermal-relic line in the scalar Rayleigh panel marks where the candidate reproduces the observed abundance, and is not an exclusion.
        Three cosmological constraints appear and should not be conflated. The
        elastic-scattering bound is computed from the Planck limit
        $\sigma < 6\times10^{-40}\,(m_\chi/\mathrm{GeV})~\mathrm{cm^{2}}$ for
        $\sigma \propto T^{2}$~\cite{Wilkinson2014DMphotonCMB,Stadler2018DMphotonPlanck}
        and lies below the halo contours by a factor $13$ at
        $m_\chi = 10^{-6}~\mathrm{GeV}$ widening to $2.3\times10^{3}$ at
        $10^{4}~\mathrm{GeV}$; no published bound exists for the $T^{4}$ and
        $T^{6}$ scalings of the Rayleigh operators, so no elastic curve is drawn
        in those panels. The Planck bound on DM annihilation to $e^{+}e^{-}$ at
        recombination~\cite{Arina2021PhotonEFT} and the CMB bound on the dipole
        moment from energy injection~\cite{Lambiase2021DMPhotonFormFactorCMB}
        both constrain the annihilation channel rather than elastic scattering
        and lie far above the halo contours; the latter is drawn dashed beyond
        $100~\mathrm{GeV}$, the upper edge of the mass prior of the published
        fit, where it is an extrapolation. The direct-detection dipole curve~\cite{PandaX4T2023,KavanaghPanciZiegler2019FaintLight} is a scaling
        estimate normalised to a published point rather than a reproduced
        exclusion, and is cut below $10~\mathrm{GeV}$ where the recoil-threshold
        turnover invalidates the asymptote.
        Two regions close the plane between them. The cyan wedge is the
        EFT-valid region, and the grey hatched region is excluded by
        thermalisation with the photon bath: below
        $\Lambda_{\rm therm} = 600$, $5$ and $320~\mathrm{GeV}$ in the three
        panels as ordered the crossed process $\gamma\gamma \leftrightarrow \chi\bar\chi$
        reaches equilibrium before nucleosynthesis, and a thermalised species
        below $0.4~\mathrm{MeV}$ is excluded~\cite{Sabti2020BBN}. Every halo
        contour lies below the wedge throughout the CDM mass range
        ($m_\chi \geq 1~\mathrm{MeV}$), and the mass at which each enters the
        wedge, $2.4\times10^{-4}$, $1.9\times10^{-5}$ and
        $1.3\times10^{-4}~\mathrm{GeV}$ respectively, lies inside the thermally
        excluded region in every case. The black $\rho^{2}$ contours are the
        $f_{\rm required}=1$ boundaries of Fig.~\ref{fig:eft_grid}; the
        $\rho^{2.5}$ contours indicate the sensitivity to the assumed halo
        profile.
    }
    \label{fig:halo_constraints}
\end{figure*}

Fig.~\ref{fig:halo_constraints} presents the $90\%$~CL sensitivity contours per
representative operator. The overlaid external constraints are digitised from
published compilations: the dipole panel from Ref.~\cite{Arina2021PhotonEFT},
collecting the LEP $Z$-decay and monojet collider limits, the XENON1T
direct-detection limit, the AMS-02, \textit{Fermi}--LAT and H.E.S.S.
indirect-detection limits and the Planck cosmological bound; the scalar Rayleigh
panel from Ref.~\cite{Barducci2025ScalarRayleigh}, giving the CTA
Galactic-centre and dwarf projections, the the LZ and XLZD direct-detection limits
and the thermal-relic line; and the dimension-7 panel from the
\textit{Fermi}--LAT and H.E.S.S. gamma-ray line searches. Reviews of the
collider and indirect-detection programmes are collected in
Refs.~\cite{Goodman2010EFT,Buchmueller2014Simplified,Kahlhoefer2017CollderReview,Boveia2018CollDMWWG,Gaskins2016IndirectReview,Slatyer2021IndirectReview}.
The headline qualitative result holds across the basis of operators that scatter
real photons (Dirac magnetic dipole, Dirac electric dipole, Dirac and Majorana
Rayleigh combining both parity variants, and scalar Rayleigh); the anapole and
charge radius set no attenuation limit at any $\Lambda$
(Sec.~\ref{subsec:eft_catalogue}).

The halo-derived contours lie in the kinematically EFT-invalid region,
$\Lambda^{2} < \max(s_{\max}, |t|_{\max})$, throughout the CDM mass range.
The CMB elastic-scattering bound does not supersede them. Planck constrains the
present-day DM--photon cross section to
$\sigma < 6\times10^{-40}\,(m_\chi/\mathrm{GeV})~\mathrm{cm^{2}}$ for
$\sigma \propto T^{2}$~\cite{Wilkinson2014DMphotonCMB,Stadler2018DMphotonPlanck},
which is the dipole case; translated onto the $(\Lambda, m_\chi)$ plane it lies
below the halo contour by a factor $13$ at $m_\chi = 10^{-6}~\mathrm{GeV}$
widening to $2.3\times10^{3}$ at $10^{4}~\mathrm{GeV}$. The reason is the
energy scaling: those constraints are set at eV-scale photon energies, where the
cross sections of the surviving operators fall below their GeV-band values by
$25$ orders of magnitude for the dipoles and by $51$ to $76$ for the Rayleigh
families. No published CMB bound exists for the $T^{4}$ and $T^{6}$ scalings the
Rayleigh operators require, so no elastic CMB curve is drawn in those panels.
The CMB bound on the dipole moment derived from the annihilation
channel~\cite{Lambiase2021DMPhotonFormFactorCMB} is a separate constraint and is
far stronger, as Sec.~\ref{subsec:uv_translation_plot} sets out. Within the
class of elastic photon--DM scattering bounds derived from diffuse Galactic
emission, the halo contours are the most stringent available; bounds derived
from the spike column around an active nucleus reach
further~\cite{FerrerHerreraIbarra2023TXS}. In none of the operator panels does
the halo-derived contour open new ground in the joint $(\Lambda, m_\chi)$ plane.

Scattering transfers momentum to the DM as well as removing photons, so the
operators considered here also perturb the DM velocity distribution. In the
early universe the photon bath is dense enough for this drag to damp
small-scale structure. That damping is the mechanism behind the CMB bounds
above, and it extends to bounds from the Milky Way satellite
population~\cite{Boehm2014Satellites,Schewtschenko2015Haloes}. In the
present-day halo the same effect is negligible. The ambient radiation field is
the CMB ($u = 0.26~\mathrm{eV\,cm^{-3}}$ at
$\omega \sim 6\times10^{-4}~\mathrm{eV}$), the dust-reprocessed infrared
($u \approx 0.3~\mathrm{eV\,cm^{-3}}$ at $\omega \sim 10^{-2}~\mathrm{eV}$)
and the stellar optical and ultraviolet field
($u \approx 0.5~\mathrm{eV\,cm^{-3}}$ at
$\omega \sim 1~\mathrm{eV}$)~\cite{MathisMezgerPanagia1983}, against
$u_\gamma \sim 10^{-7}~\mathrm{eV\,cm^{-3}}$ in the gamma band. The
momentum-transfer rate is the integral of the photon flux against the cross
section, and $\sigma \propto \omega^{k}$ with $k \geq 2$ makes the gamma rays
dominate that integral despite their small share of the energy density. Even
for the shallowest case, the dipole at $k = 2$, the optical contribution is
suppressed relative to the GeV band by
$(\omega_{\rm opt}/\omega_{\rm GeV})^{2}\,u_{\rm opt}/u_\gamma \sim 10^{-11}$,
the infrared by $10^{-16}$ and the CMB by $10^{-19}$. The DM kinetic energy
density is of order $10^{2}~\mathrm{eV\,cm^{-3}}$, so even if every gamma ray
crossing the halo scattered depositing its entire energy, the DM would gain
less than one part in $10^{3}$ of its kinetic energy over a Hubble time. At the
couplings probed in this work, the transfer is many orders of magnitude
smaller, since at most a fraction $\tau \sim 10^{-2}$ of the photons scatter,
with scatters depositing only a fraction $\Delta\omega/\omega$ of the photon
energy.

The low-mass window in which the contours do become EFT-consistent requires a
separate cosmological assessment. On the elastic
plane the comparison above is quantitatively sound: converting to CMB-epoch
energies, the dipole at $\Lambda = 0.29~\mathrm{GeV}$ gives a present-day cross
section $\sigma \simeq 2\times10^{-49}~\mathrm{cm^{2}}$, some six orders of
magnitude below the $\sigma \propto T^{2}$ bound at $m_\chi =
240~\mathrm{keV}$~\cite{Wilkinson2014DMphotonCMB}, and the Rayleigh operators
fall short of the strongest constant-cross-section
bounds~\cite{Stadler2018DMphotonPlanck,Escudero2018gammaCDM} by 15--26 orders of magnitude. Elastic scattering is not, however, the binding
early-universe probe. By crossing, the same operators drive $\gamma\gamma
\leftrightarrow \chi\bar\chi$, and at $T = 1~\mathrm{MeV}$ that rate exceeds the
Hubble rate by roughly $10^{12}$ for the dimension-5 and dimension-6 windows and
by $10^{3}$ for dimension-7, a computation that lies inside the EFT validity
range since $s \sim T^{2} \ll \Lambda^{2}$. A species of mass
$12$--$240~\mathrm{keV}$ in thermal contact with the photon bath is excluded:
BBN alone requires $m_\chi > 0.4~\mathrm{MeV}$ for any
thermal relic in equilibrium with the plasma~\cite{Sabti2020BBN}, and
annihilation to photons after neutrino decoupling would drive $N_{\rm eff}$ to
$0.8$ for the Dirac dipole or $1.8$ for the real scalar, against the measured
$2.99 \pm 0.17$~\cite{Planck2018CosmoParams,BoehmDolanMcCabe2013}. The
dimension-7 case instead freezes out while relativistic, adding $\Delta N_{\rm
eff} \simeq 1$ and a relic density two to three orders of magnitude above the
observed one. Avoiding thermal contact would require $\Lambda \gtrsim
600$, $320$ and $5~\mathrm{GeV}$ for the dimension-5, 6 and
7 operators respectively, far above the reach of this observable at any exposure
considered here. The low-mass contours are therefore internal consistency
statements of the attenuation analysis rather than exclusions of open parameter
space, since a particle carrying these couplings at these masses is already
forbidden as DM in any standard thermal history. Stellar and
SN1987A energy-loss bounds~\cite{Chu2019Stellar} do not bite at these couplings,
the dark states being trapped rather than free-streaming, so cosmology provides
the decisive test.

A statement sharper than EFT-invalidity is available for the dipoles: the excluded scales correspond to a
dipole moment that runs from $\mu = 2c_M/\Lambda \simeq 6~\mathrm{GeV}^{-1}$ at
$m_\chi = 60~\mathrm{GeV}$ to $7.5~\mathrm{GeV}^{-1}$ at $m_\chi \simeq 650~\mathrm{GeV}$, generated
radiatively by a charged mediator of mass $M$ and coupling $g$, as in the
one-loop estimate of Eq.~\eqref{eq:ew_doublet_dipole},
$\mu \sim e g^{2}/(64\pi^{2} M)$. This places $M$ between
$\mathcal{O}(0.1)~\mathrm{MeV}$ at $g \sim 1$ and $\mathcal{O}(10)~\mathrm{MeV}$
at the perturbative ceiling $g^{2} = 16\pi^{2}$. Direct searches at LEP require
any new charged state above roughly
$100~\mathrm{GeV}$~\cite{LEP2004ChargedParticles,FortinTait2012}, four to six
orders of magnitude above that window. The region these contours exclude
therefore contains no perturbative UV completion, which is the same
conclusion reached independently for the Higgs portal and for the gravitational
channel in Sec.~\ref{sec:UV_revisit}.

The contours do, however, extend the covered mass range. Published curves for
these operators are plotted only up to $m_\chi = 10^{5}~\mathrm{GeV}$ for the
dipoles, charge radius and anapole~\cite{Arina2021PhotonEFT} and
$10^{4}~\mathrm{GeV}$ for the Rayleigh
operators~\cite{KavanaghPanciZiegler2019FaintLight}. Recasting the H.E.S.S.
gamma-line search~\cite{HESS2013GammaLines} onto the Rayleigh basis ourselves
carries that comparison to the $2.5\times10^{4}~\mathrm{GeV}$ upper edge of its
quoted mass range, which is as far as any existing measurement reaches. The halo
scan returns a $90\%$~CL contour out to $10^{8}~\mathrm{GeV}$ for the dipoles and
the dimension-7 Rayleigh operators, and to $3.7\times10^{5}~\mathrm{GeV}$ for the
scalar Rayleigh. Above $10^{5}~\mathrm{GeV}$ no published constraint on
photon--DM scattering through these operators exists from collider,
direct-detection, indirect-detection or cosmological data. Two qualifications
bound what that coverage means. Existing curves stop at
$10^{4}$--$10^{5}~\mathrm{GeV}$ because partial-wave unitarity already excludes
a thermal relic above that mass~\cite{Griest1990Unitarity}, not because those
probes lose sensitivity there. Additionally, the region lies deep inside the EFT-invalid
wedge, and above $m_\chi \simeq 1.1~\mathrm{TeV}$ lies above the closure of the
energy-migration channel, so what the contour constrains there is removal by
angular deflection out of the region of interest rather than a spectral
distortion or a cutoff scale.

The reshaping observable defined in Eq.~\eqref{eq:atten+reshape} produces no
$90\%$~CL contour for the dipoles or the fermionic Rayleigh operators. This is
not in tension with the attenuation limits of Sec.~\ref{sec:eft_operators}, and
the two exclusion regions are not nested. Every point the reshaping observable
would exclude requires $\tau \gtrsim 0.35$, whereas the single-scatter
truncation of Eq.~\eqref{eq:T_app} is quantitatively reliable only for
$\tau \lesssim 0.3$, where the neglected $\mathcal{O}(\tau^{2}/2)$ term stays
below $5\%$ of the transported flux. That figure understates the error on the
quantity actually fitted. The reshaping signal is a near-cancellation between
the attenuation and in-scatter terms, so a $5\%$ error on the total is an
$\mathcal{O}(1)$ error on the residual, and the ceiling should be read as
permissive rather than conservative. The excluded and valid regions abut rather than overlap, and the
dipole case is degenerate with the ceiling to within the resolution of the
scan. The scalar Rayleigh alone retains a contour, at
$\Lambda \simeq 0.19$--$0.39~\mathrm{GeV}$ over
$m_\chi \simeq 7\times10^{-5}$--$1.6\times10^{-3}~\mathrm{GeV}$, weaker than the
attenuation limit and therefore not binding.

Equation~\eqref{eq:atten+reshape} retains the survival term
$\Phi_{\rm src}\,e^{-\tau}$ that drives the attenuation scan, but adds the in-scatter
term $\sum_j K_{ij}\,\tau_j\,\Phi_{{\rm src},j}\,e^{-\tau_j}$, which returns to each
energy bin the photons the survival term removes. That refill is substantial at
low mass and negligible at high: the net spectral distortion is $0.15$ of the
bare $1-e^{-\tau}$ suppression at $m_\chi \simeq 10^{-3}~\mathrm{GeV}$ and $0.78$
at $10^{3}~\mathrm{GeV}$ for the dipoles, tracking the fraction of scattered
photons that remain within the region of interest. The cancellation is
therefore real, but it is not what removes the contour. Because
$\tau \ll 1$ throughout the spectral window, both the survival deficit and the
in-scatter refill are $\mathcal{O}(\tau)$. It is their near-cancellation, not
the smallness of $\tau$ alone, that sets the size of the reshaping signal. What
limits the arm is not that signal but the domain over which it can be computed:
the optical depth needed to reach $\Delta\chi^{2} = 4.61$ is larger than the
optical depth at which a one-scatter treatment remains reliable. Extending it
therefore requires a resummed transfer including multiple scattering rather than
a larger exposure, and we do not attempt that here.

These results are not a failure of the analysis. They quantify how far below the
\textit{Fermi}--LAT spectral-detection threshold the optical depth lies for any
natural Wilson coefficient, and the value of the result is methodological.

Existing photon--DM elastic-scattering results fall into three classes.
CMB-epoch analyses constrain the scattering cross section at
recombination~\cite{Wilkinson2014DMphotonCMB,Stadler2018DMphotonPlanck}.
Late-universe gamma-ray studies of individual sources report a single effective
$\sigma_{\text{DM}\gamma}$ at fixed $m_\chi$, either as an upper
limit~\cite{FerrerHerreraIbarra2023TXS} or, for NGC 1068, as a claimed
indication of absorption~\cite{Herrera2025NGC1068}. In both cases the constraint
is on one cross-section normalisation rather than on an operator basis with a
predicted energy dependence. The remaining line-of-sight treatment is tied to
specific UV completions~\cite{acar2026darkmatterredblue}. The present work is the
intersection of these with the operator-resolved EFT treatments of the
annihilation channel~\cite{KavanaghPanciZiegler2019FaintLight,%
Barducci2025ScalarRayleigh}: a basis spanning dimension-5 to dimension-7, each
member with its own predicted energy dependence, constrained by the elastic
signature in a late-universe astrophysical gamma-ray dataset.

As shown in Sec.~\ref{subsec:cross_dataset}, the framework transfers directly to
other diffuse-emission datasets and to next-generation gamma-ray instruments.
The direction of the gain is set by a competition. The validity wedge scales as
$\Lambda \propto \sqrt{\omega_{\max}}$ at low mass
(Sec.~\ref{subsec:eft_criteria}), while the exclusion contour scales as
$\Lambda_{90} \propto \omega_{\max}^{\,k/p}$, with $k$ the energy index of the
cross section and $p = 4$ ($6$) the power of $\Lambda$ for the dipole and scalar
(dimension-7) operators. The wedge--contour gap therefore closes as
$(\tfrac{1}{2} - k/p)$ per decade of $\omega_{\max}$: it is scale-invariant for
the dipoles ($k/p = 1/2$) and closes with increasing $\omega_{\max}$ for
every Rayleigh operator ($k/p = 2/3$--$1$). TeV instruments such as CTA and
SWGO~\cite{CTAConsortium2019Science,SWGOScience2025} are therefore the
favourable direction for the Rayleigh family, subject to the self-consistency
requirement $\omega_{\max} \lesssim \Lambda$; for the dipoles no choice of band
closes the gap, and only improved fractional precision does.

%------------------------------------------------------------------
\subsection{Cross-dataset consistency}
\label{subsec:cross_dataset}
%------------------------------------------------------------------

\begin{figure*}[t]
    \centering
    \includegraphics[width=\textwidth]{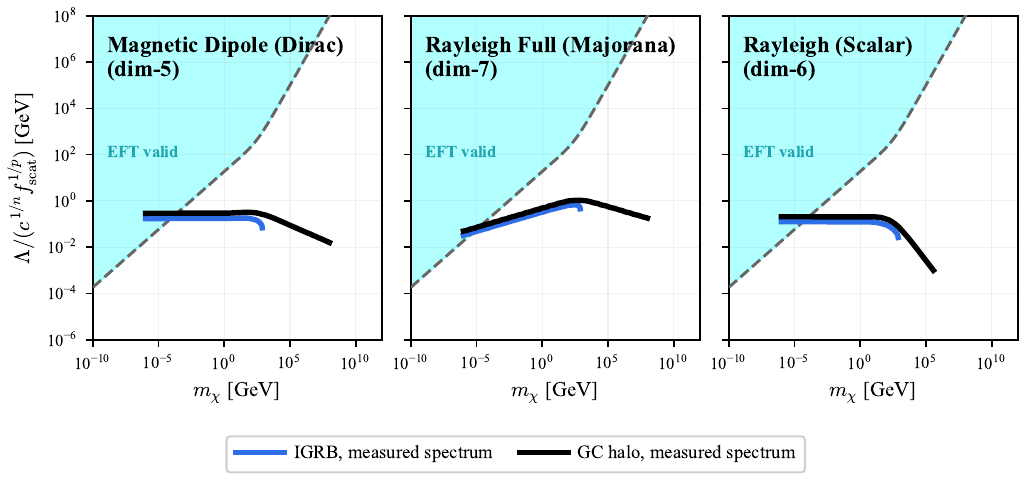}
    \caption{
Multi-dataset $90\%$~CL sensitivity contours for the same three representative
operators, on the same axes, as Fig.~\ref{fig:halo_constraints}. The IGRB curve is not an independent limit on the operators.
It is a consistency test of the halo result against a dataset with different
foregrounds and a different removal mechanism, and it is shown only for that
purpose. Colour encodes the data source: black is the
\textit{Fermi}--LAT Galactic-centre halo
posterior~\cite{StenhouseGhagDeppisch2026Totani}, blue the
\textit{Fermi}--LAT isotropic diffuse gamma-ray
background~\cite{Ackermann2015IGRB}. Both take the measured spectrum as the
intrinsic pre-scattering template. Throughout
the displayed CDM mass range all contours lie in the EFT-invalid region
below the cyan-shaded validity wedge.
    }
    \label{fig:multi_dataset_overlay}
\end{figure*}

The framework of Sec.~\ref{subsec:halo_dataset}--\ref{subsec:halo_results} is not tied to the Galactic-centre halo posterior. Every ingredient (the test statistic of Eq.~\eqref{eq:delta_chi2_general}, the transfer matrix of Eq.~\eqref{eq:atten+reshape}, and the EFT operator basis of Sec.~\ref{sec:eft_operators}) transfers to any \textit{Fermi}--LAT dataset that provides a per-bin photon flux with a well-characterised statistical uncertainty and a line-of-sight column density. We use additional datasets as an independent cross-check.

\paragraph{Isotropic diffuse background (IGRB).} We use the Ref.~\cite{Ackermann2015IGRB} measurement of the isotropic diffuse gamma-ray flux (their Foreground Model A, 26 bins spanning $100~\mathrm{MeV}$--$820~\mathrm{GeV}$) as the reference spectrum, truncated at $200~\mathrm{GeV}$ as for the halo posterior (Sec.~\ref{subsec:eft_criteria}), which gives $\omega_{\max} = 172.2~\mathrm{GeV}$ for this dataset, with the cosmological baseline column density $J_{\rm cosmo} = 1.37\times10^{22}~\mathrm{GeV\,cm^{-2}}$ of Eq.~\eqref{eq:cosmo_baseline}. The test statistic and boundary extraction are identical to the halo pipeline, with the ROI-integrated column of Sec.~\ref{subsec:halo_dataset} replaced by the scalar $J_{\rm cosmo}$.

\paragraph{Dwarf spheroidal galaxies (dSphs).} A parallel dSph analysis using the McDaniel~2024 legacy release~\cite{McDaniel2024DsphLegacy}, stacking the eight classical dwarfs with linear column densities computed from the Geringer-Sameth, Koushiappas \& Walker~2015 dwarf density profiles~\cite{GeringerSameth2015Dfactors}, gives a stacked scattering column $J_{\rm dSph}^{\rm stack} \approx 1.9\times10^{22}~\mathrm{GeV\,cm^{-2}}$ that is comparable to $J_{\rm cosmo}$ and within a factor of $\sim 2.5$ of $J_{\rm ROI}$. At current \textit{Fermi}--LAT dSph exposure the per-bin SED significance leaves the normalisation-profiled fit of Eq.~\eqref{eq:halo_chi2} degenerate: any predicted scattering distortion is absorbed by driving the intrinsic source normalisation toward zero. Setting a physical dSph scattering bound therefore requires a fixed-normalisation two-component analysis with an external dSph annihilation cross-section upper limit as an anchor, which we leave to a follow-up study.

Fig.~\ref{fig:multi_dataset_overlay} repeats Fig.~\ref{fig:halo_constraints} with the external bounds stripped away, leaving the halo bound and its IGRB cross-check on a common $(\Lambda, m_\chi)$ plane so that their agreement can be read off directly. We do not present the IGRB curve as a second limit on the operator basis. It is a test of whether the halo result survives a change of dataset, foreground treatment and removal mechanism, and it is the agreement rather than the curve that carries the information.
The two agree to within a factor of $1.4$ across every operator below $m_\chi \sim 100~\mathrm{GeV}$, and are flat in mass there, consistent with the two line-of-sight columns, $J_{\rm ROI}$ and $J_{\rm cosmo}$, agreeing to within a factor of a few. For an isotropic source against a uniform cosmological scatterer, out-scattering from any line of sight is compensated exactly by in-scattering from the surrounding sky, so energy migration is the only channel by which a photon leaves its bin. That channel closes near $m_\chi \simeq 1.1~\mathrm{TeV}$, where even a full backscatter shifts a photon by less than its bin width, and the IGRB contour terminates there. The ratio climbs as that closure is approached, because the IGRB bound weakens while the halo bound, which retains angular deflection out of the region of interest, does not. The halo and IGRB arms therefore constrain through different mechanisms, making their agreement a test rather than a repetition.
The small residual offset tracks the different spectral shapes and statistical uncertainty profiles of the two measurements.
We also repeat the fit with an annihilation template in place of the measured spectrum, using the PPPC tabulation~\cite{Cirelli2011PPPC} at the best-fit annihilator masses of Ref.~\cite{StenhouseGhagDeppisch2026Totani}, $m_\chi^{\rm ann} \simeq 0.56~\mathrm{TeV}$ ($W^{+}W^{-}$) and $0.72~\mathrm{TeV}$ ($b\bar b$), which fixes the annihilator and leaves the scatterer mass $m_\chi^{\rm scat}$ as the only free mass in the fit.
The variant places its contour where the measured-template analysis would
suggest, but the statistic that produces it is not an upper limit, and the two
facts should be kept apart. Both channels place their contour within a factor of a few of the
measured-template dipole contour computed on the same numerical generation. We
quote the ratio rather than an absolute scale, since this variant has not been
regenerated on the removal path used for the quoted limits. On that evidence, the choice of source
ansatz shifts the exclusion by a factor of a few rather than reversing the
qualitative conclusion. The fit behind it, however, is poor:
$\chi^{2}/\mathrm{dof} \simeq 9$ over the retained band, and the $\chi^{2}$
minimum lies at an interior point with $\tau > 0$, the fit preferring some
scattering to none by $\Delta\chi^{2} \simeq 10$. Its $90\%$~CL region is
therefore a confidence interval around a preferred non-zero optical depth rather
than a bound, and we do not draw it in
Fig.~\ref{fig:multi_dataset_overlay} or treat it as an independent constraint.
All bounds sit inside the EFT-invalid region across the operator basis
throughout the CDM mass range, and the apparent extension of coverage above
$m_\chi \simeq 10^{5}~\mathrm{GeV}$ carries the qualifications of
Sec.~\ref{subsec:halo_results}.

These comparisons address one of two systematic concerns and leave the other
open. The first is settled: the halo-derived bound is not an artefact of the
pixel-level halo-fitting pipeline of
Ref.~\cite{StenhouseGhagDeppisch2026Totani}, since an entirely independent
dataset analysed with a wholly different foreground-modelling procedure
reproduces the same $\Lambda$ scale of exclusion. The second is addressed in position but not in rigour. The annihilation
template puts its contour within a factor of a few of the measured-template
result, so the conclusion does not appear to depend on using the data as its own
source model, but the underlying fit is poor enough that we do not claim the
source-model dependence as tested. The price
throughout is that the bounds remain non-competitive with existing channels in
the EFT-invalid region.

%==================================================================
\section{UV-Complete Translations}
\label{sec:uv_translations}
%==================================================================
\begin{figure*}[t]
    \centering
    \includegraphics[width=\textwidth]{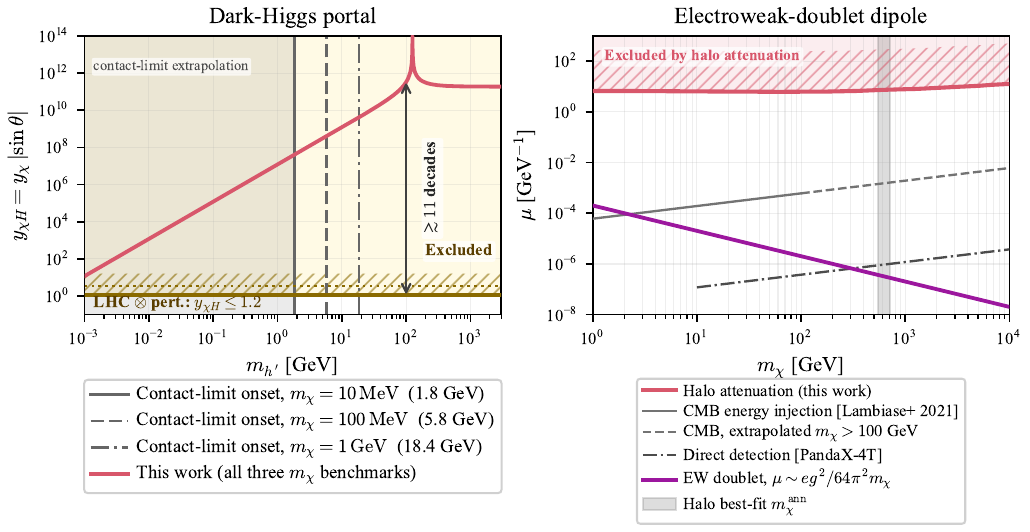}
    \caption{
        UV-completion parameter-space translations of the halo-derived photon--DM
        scattering sensitivity. Left: dark-Higgs portal on the
        $(m_{h\p}, y_{\chi H})$ plane via Eq.~\eqref{eq:dh_to_rayleigh}. The
        three benchmark scatterer masses give one common curve, since
        $y_{\chi H}$ absorbs $m_\chi/v\p$ and the halo bound is flat in mass;
        they differ only in where the contact limit fails (grey, one onset each
        at $1.8$, $5.8$ and $18.4~\mathrm{GeV}$). The curve diverges at
        $m_{h\p} = m_h$, where the two propagator contributions cancel, and
        saturates above it. The solid gold line is the combined LHC mixing and
        perturbativity ceiling $y_{\chi H} \leq 1.2$, which the required
        coupling exceeds everywhere, and the dotted gold line above it is the
        dark-sector perturbativity bound $y_{\chi H} = \sqrt{4\pi}$ taken alone. Right: electroweak-doublet dipole portal on
        the $(m_\chi, \mu)$ plane, with the halo bound (pink), the one-loop
        doublet prediction of Eq.~\eqref{eq:ew_doublet_dipole} (purple) and the
        best-fit annihilator band of
        Ref.~\cite{StenhouseGhagDeppisch2026Totani}. The CMB
        energy-injection~\cite{Lambiase2021DMPhotonFormFactorCMB} and
        direct-detection~\cite{PandaX4T2023,KavanaghPanciZiegler2019FaintLight}
        bounds are overlaid as scaling estimates, dashed where extrapolated.
        Both are stronger than the attenuation bound for unsplit Dirac DM, and
        both close once the doublet is split, which is the regime this panel is
        about (Sec.~\ref{subsec:uv_translation_plot}). No panel is shown for the
        kinetically mixed dark photon, whose matched operators have identically
        vanishing real-photon amplitudes
        (Sec.~\ref{subsec:darkphoton_to_dim56}).
    }
    \label{fig:uv_translation}
\end{figure*}
The EFT limits of Sec.~\ref{sec:halo_limits} are stated in terms of $(\Lambda, m_\chi)$ at a fixed benchmark Wilson coefficient. Ref.~\cite{StenhouseGhagDeppisch2026Totani} fixes a candidate model space for the halo-like excess: an annihilator at $m_\chi^{\rm ann} \simeq 0.56$--$0.72~\mathrm{TeV}$, resonantly boosted by a velocity-dependent enhancement, whose Sommerfeld realisation admits mediators from the sub-GeV range up to $\mathcal{O}(10)~\mathrm{GeV}$. At least five explicit realisations have been published~\cite{Yamashita2026TotaniHiggsPortal,Jho2025TotaniDilaton,NomuraTotani2026Higgsino,Yoshimatsu2026SecludedWW,Murayama2025BreitWigner}, spanning sub-GeV mediators~\cite{Yamashita2026TotaniHiggsPortal}, a $\sim 12$~GeV scalar cascade~\cite{Jho2025TotaniDilaton}, a resonant Breit-Wigner realisation~\cite{Murayama2025BreitWigner}, and an electroweak-doublet case that is naturally inelastic, with a splitting set by the halo kinetic energy~\cite{NomuraTotani2026Higgsino}. This section asks which members of that model space photon--DM scattering can probe, and through which operators. We translate the halo-derived bounds onto three illustrative completions, taken in turn.

%------------------------------------------------------------------
\subsection{Dark-Higgs portal $\to$ Rayleigh operator}
\label{subsec:dark_higgs_to_rayleigh}
%------------------------------------------------------------------

The dark-Higgs portal of Sec.~\ref{subsec:dark_higgs} generates a contact $\gamma\gamma$--$\chi\bar\chi$ four-point amplitude that maps directly onto the fermionic Rayleigh operator $\mathcal{O}_{R}^{M}$ of Eq.~\eqref{eq:O_rayleigh_majorana}. The effective Wilson coefficient is
\begin{equation}
\frac{c_s}{\Lambda^{3}} \;\sim\; \frac{\alpha}{\pi v}\,y_{\chi H}\cos\theta\,\mathcal{F}_{\rm loop}\left(\frac{1}{m_{h}^{2}}-\frac{1}{m_{h\p}^{2}}\right),
\label{eq:dh_to_rayleigh}
\end{equation}
where $y_{\chi H} = (m_\chi/v\p)\sin\theta$ is the effective $\bar\chi\chi h$ coupling defined in Eq.~\eqref{eq:y_chi_H} and $\mathcal{F}_{\rm loop}$ collects the $W$- and top-quark loop form factors of Eq.~\eqref{eq:dsdo_revisit}. Both propagators appear because $h$ and $h\p$ each mediate the $\gamma\gamma$--$\chi\bar\chi$ transition, so the contact limit requires the momentum transfer to be small compared with either mass, $|t| \ll m_{h}^{2},\,m_{h\p}^{2}$, rather than the photon energy to be small. The two contributions enter with opposite sign: photons couple to the doublet component $H$ and the scatterer to the singlet $H\p$, so the mixing angle appears as $\sin\theta\cos\theta$ in both terms while the projections onto the two mass eigenstates carry opposite signs. The combination therefore vanishes in the degenerate limit $m_{h\p} \to m_{h}$, as it must, since two mass-degenerate states can be rotated into one another and the mixing angle is then unphysical. In the two asymptotic regimes $m_{h\p} \ll m_{h}$ and $m_{h\p} \gg m_{h}$ the bracket reduces to $-1/m_{h\p}^{2}$ and $+1/m_{h}^{2}$ respectively, so the magnitude of the matching is unaffected there and only the region of near-degeneracy is sensitive to the relative sign. With $|t|_{\max} \simeq 2 m_\chi \omega_{\max}$ for a light scatterer this is satisfied for $h$ throughout, and for $h\p$ above $m_{h\p} \gtrsim \sqrt{2 m_\chi \omega_{\max}}$, which is $1.8$, $5.8$ and $18.4~\mathrm{GeV}$ at the three benchmark scatterer masses of Fig.~\ref{fig:uv_translation}. Substituting Eq.~\eqref{eq:dh_to_rayleigh} into the halo-derived $\Lambda_R(m_\chi)$ exclusion of Sec.~\ref{subsec:halo_results} maps the limit onto the dark-Higgs parameter space $(\sin\theta, v\p, m_\chi)$, shown in the left panel of Fig.~\ref{fig:uv_translation}. The combined ceiling $y_{\chi H} \lesssim 1.2$, set jointly by the LHC mixing-angle bound $\sin\theta \lesssim 0.33$ of Eq.~\eqref{eq:mixing_bound} and perturbativity, is overlaid for comparison. As expected from Sec.~\ref{subsec:mchi_tau_plot}, the scattering limit is many orders of magnitude weaker than the LHC. The value of the translation is not in the bound, but in the channel: a route from astrophysical photon-attenuation data to dark-Higgs parameter space that is structurally independent of every other constraint on the same plane. The dark Higgs $h\p$ is itself a concrete candidate for the light mediator the halo fit demands: a light scalar mediator is the construction of Ref.~\cite{Jho2025TotaniDilaton}, there realised at $m_\varphi \sim 12$~GeV rather than in the sub-GeV window, so the Rayleigh translation of this section is a probe of the mediator sector of the halo-like excess model space.

%------------------------------------------------------------------
\subsection{Kinetically mixed dark photon: a structural blind spot}
\label{subsec:darkphoton_to_dim56}
%------------------------------------------------------------------

The kinetically mixed dark photon~\cite{Holdom1986,Pospelov2008Secluded,BauerFoldenauerJaeckel2018} is the most natural sub-GeV portal for the halo-preferred mediator. A dark $U(1)\p$ gauge boson $A\p$ couples to the SM only through kinetic mixing,
\begin{equation}
    \mathcal{L} \supset
    -\tfrac{1}{4}F\p_{\mu\nu}F\p^{\mu\nu}
    + \tfrac{1}{2}M_{A\p}^{2}A\p_\mu A\p^{\mu}
    - \tfrac{\epsilon}{2}F_{\mu\nu}F\p^{\mu\nu}
    + g_\chi\,\bar\chi\gamma^{\mu}\chi\,A\p_\mu ,
    \label{eq:kinetic_mixing}
\end{equation}
so that after diagonalising the kinetic term the dark current reaches the photon only through $A\p$ exchange. It matches directly onto the dimension-6 charge-radius operator of Sec.~\ref{subsec:eft_catalogue}, and onto the anapole if the dark current carries an axial component~\cite{AriasGamboaTapia2019Anapole}.
Because those operators couple only through the form factor $\epsilon\,g_\chi\,q^{2}/(q^{2}-M_{A\p}^{2})$, which vanishes for an on-shell photon ($q^{2}=0$), their real-photon amplitude is exactly zero and halo attenuation places no bound on $(\epsilon, M_{A\p})$ at any exposure.
The testable plane belongs instead to accelerator probes that access the virtual-photon coupling directly, chiefly BaBar~\cite{BaBar2017DarkPhoton} and NA64~\cite{NA64Andreev2023}.
The corollary is what makes this null result useful for the halo interpretation of Ref.~\cite{StenhouseGhagDeppisch2026Totani}:
if the halo-like excess dark sector mediator is a pure kinetically mixed dark photon,
the light dark sector produces no attenuation signature at all,
and any future detection of operator-induced attenuation in this mass range would instead point to a dipole- or Rayleigh-type photon coupling.

%------------------------------------------------------------------
\subsection{Electroweak-doublet dipole portal}
\label{subsec:ew_doublet_dipole}
%------------------------------------------------------------------

The natural UV completion for the operators attenuation can see is one in which the DM acquires a dipole moment from a charged loop.
The minimal realisation connected to the halo interpretation is a Dirac electroweak doublet, the higgsino-like candidate class already proposed for the 20~GeV halo-like excess~\cite{NomuraTotani2026Higgsino}.
After electroweak symmetry breaking, $W$--chargino loops endow the neutral Dirac state with a magnetic dipole moment of parametric size~\cite{HambyeXu2021Dipoles}
\begin{equation}
    \mu \;\sim\; \frac{e\,g^{2}}{64\pi^{2}\,m_\chi}\,\mathcal{F}\!\left(\tfrac{m_W}{m_\chi}\right),
    \qquad \mathcal{F} = \mathcal{O}(1),
    \label{eq:ew_doublet_dipole}
\end{equation}
so the annihilator itself carries the dimension-5 operator of Eq.~\eqref{eq:O_magneticdipole_dirac} with $\mu = 2c_M/\Lambda$.
Because the dipole cross section depends only on this combination (Eq.~\eqref{eq:dipole_compton_m2}), the halo boundary $\Lambda_{90}(\mchi)$ of Fig.~\ref{fig:halo_constraints} converts directly to an excluded dipole moment $\mu > 2/\Lambda_{90}(\mchi)$.
This translation carries a discriminant specific to the halo-like excess. The direct detection escape mechanism invoked for the higgsino-like candidate, an $\mathcal{O}(100~\mathrm{keV})$ inelastic splitting~\cite{NomuraTotani2026Higgsino},
splits the Dirac state into two Majorana states $\chi_{1,2}$. The diagonal moments then vanish ($\bar\chi_i\sigma^{\mu\nu}\chi_i \equiv 0$),
but the dipole is not destroyed.
It survives as a transition moment $\bar\chi_2\sigma^{\mu\nu}\chi_1 F_{\mu\nu}$,
the magnetic inelastic DM structure of Ref.~\cite{WeinerYavin2012RayDM}.
The two probes respond to this structure in sharply different ways. Direct detection through the transition moment is kinematically gated: exciting $\chi_1 N \to \chi_2 N$ costs the splitting $\delta$, and for $\delta$ above the $\mathcal{O}(10\mbox{--}100~\mathrm{keV})$ kinetic-energy budget of halo DM the channel closes. That closure is the escape.
Photon attenuation carries no such gate:
the two-insertion Compton amplitude passes through a virtual $\chi_2$ and returns to the ground state, and for $E_\gamma \sim \mathrm{GeV} \gg \delta$, the elastic cross section is indistinguishable from the unsplit Dirac case.
The dipole bound of this section therefore applies to the split and unsplit
realisations alike. That indifference to $\delta$ is what gives it value, because
the two stronger constraints on the same plane do not share it, and both switch
off for the same reason the model invokes the splitting in the first place.
Direct detection closes kinematically: writing $\mu_{\chi N}$ for the DM--nucleus
reduced mass, the energy available to upscatter a $0.7~\mathrm{TeV}$ scatterer on
xenon is $\tfrac{1}{2}\mu_{\chi N} v^{2} \simeq 31~\mathrm{keV}$, below the
$\mathcal{O}(100)~\mathrm{keV}$ splitting. The CMB annihilation bound closes
thermally: annihilation through a transition moment requires the excited state,
whose population at recombination is suppressed by $e^{-\delta/T}$ with
$\delta/T \sim 4\times10^{5}$. Attenuation is the only one of the three that
survives, and Sec.~\ref{subsec:uv_translation_plot} quantifies all three and
shows what that implies for the viability of the candidate.

%------------------------------------------------------------------
\subsection{UV-completion parameter-space bounds}
\label{subsec:uv_translation_plot}
%------------------------------------------------------------------
Fig.~\ref{fig:uv_translation} translates the halo-derived $(\Lambda, m_\chi)$ exclusions of Fig.~\ref{fig:halo_constraints} onto the native parameter spaces of the two completions of this section that inherit genuine attenuation observables: the dark-Higgs portal (left) and the electroweak-doublet dipole (right). No panel exists for the kinetically mixed dark photon, as the matched operators have vanishing real-photon amplitude and halo attenuation places no bound on the $(\epsilon, M_{A\p})$ plane (Sec.~\ref{subsec:darkphoton_to_dim56}).

In the left panel, motivated by the analysis of Ref.~\cite{StenhouseGhagDeppisch2026Totani}, in which the halo-like excess is best fit by a heavy annihilator ($m_\chi^{\rm ann} \simeq 0.56$--$0.72~\mathrm{TeV}$) resonantly boosted by a sub-GeV mediator, the scatterer is the light component of the two-component dark sector. We map the fermionic-Rayleigh halo bound onto the dark-Higgs portal coupling $y_{\chi H} = y_\chi\sin\theta$ as a function of $m_{h\p}$ using Eq.~\eqref{eq:dh_to_rayleigh}, at three benchmarks bracketing the preferred window, $m_\chi \in \{10, 100, 1000\}~\mathrm{MeV}$. Because $y_{\chi H}$ absorbs the factor $m_\chi/v\p$, and because the halo bound on $\Lambda_R$ is itself flat in mass across this window, the three benchmarks give a single required-coupling curve, distinguished only by the mediator mass below which the contact limit fails. That curve rises with $m_{h\p}$, diverges at $m_{h\p} = m_h$ where the two propagator contributions cancel exactly, and saturates above the degeneracy at the value set by the SM Higgs propagator alone. Throughout the region in which the contact limit holds it sits at least eleven orders of magnitude above the maximum achievable portal coupling, set jointly by the LHC signal-strength bound $|\sin\theta| \lesssim 0.33$~\cite{ATLASHiggs2022,CMSHiggs2022} and the perturbativity bound $y_\chi \leq \sqrt{4\pi}$ (together $y_{\chi H} \lesssim 1.2$). The curve approaches the ceiling only at the extreme low-$m_{h\p}$ end of the displayed range, which lies inside the shaded contact-limit extrapolation and is not a valid matching there. This is consistent with the direct UV analysis of Sec.~\ref{sec:UV_revisit}. Producing a measurable Rayleigh coefficient through the dark-Higgs portal requires a non-perturbative or collider-excluded portal coupling, independently of whether the scatterer is heavy or light.

The right panel maps three bounds on the $(m_\chi, \mu)$ plane against the
one-loop electroweak-doublet prediction of
Eq.~\eqref{eq:ew_doublet_dipole}. At the centre of the best-fit band of
Ref.~\cite{StenhouseGhagDeppisch2026Totani} ($m_\chi^{\rm ann} \simeq
0.56$--$0.72~\mathrm{TeV}$, shaded) the halo attenuation bound sits at
$\mu \lesssim 7.5~\mathrm{GeV}^{-1}$, the CMB energy-injection bound at
$1.5\times10^{-3}~\mathrm{GeV}^{-1}$ and the direct-detection bound at
$9.5\times10^{-7}~\mathrm{GeV}^{-1}$, against a predicted
$3.2\times10^{-7}~\mathrm{GeV}^{-1}$. On the unsplit plane the attenuation bound
is therefore not competitive, trailing the other two by $3.7$ and $6.9$ dex,
and since $\Delta\chi^{2} \propto \sigma^{2} \propto \mu^{8}$ in the small-$\tau$
regime no realistic exposure closes that gap.

The informative comparison is a different one. Direct detection sits within a
factor of $2.9$ of the predicted doublet moment, so the unsplit realisation of
this candidate is close to being excluded already, and the inelastic splitting is
invoked in this model class for exactly that purpose. The splitting does not,
however, relieve the AMS-02 antiproton bound of
Ref.~\cite{WangDuan2026AntiprotonExclusion} on $W^{+}W^{-}$ annihilation, which
constrains the ground-state annihilation rate directly; what follows is therefore
a statement about the complementarity of the three photon-sector probes on this
plane, not a claim that the candidate survives all constraints. The splitting is thus not an incidental feature of the construction but
the reason it survives, and the surviving corner is precisely where the two
leading constraints switch off: direct detection kinematically, once the
available recoil energy falls below $\delta$, and the CMB thermally, once
annihilation through a transition moment is Boltzmann-suppressed at
recombination. Photon attenuation is the only one of the three that does not
care (Sec.~\ref{subsec:ew_doublet_dipole}). The complementarity is therefore
kinematic rather than algebraic, and it is sharpest exactly where the model is
still alive.

Two of the three curves are overlaid as scaling estimates rather than
reproductions of the published exclusions, each normalised to a quoted bound and
extended as $\mu \propto \sqrt{m_\chi}$. That common slope is not a coincidence
of parametrisation: both are rate limits on a cross section proportional to
$\mu^{2}$, carrying one factor of the number density $\rho/m_\chi$ at fixed mass
density, so both give $\mu^{2} \propto m_\chi$. The direct-detection curve is
cut below $10~\mathrm{GeV}$, where the recoil-threshold turnover invalidates the
asymptote.

Taken together, the three translations of this section specify a programme rather than a single result. Mediator sectors built on a scalar, Higgs-like portal are probed through the Rayleigh operators (Sec.~\ref{subsec:dark_higgs_to_rayleigh}), mediators that couple through kinetic mixing alone are unprobeable by attenuation at any exposure and are left entirely to accelerator searches (Sec.~\ref{subsec:darkphoton_to_dim56}), and electroweak-multiplet annihilators are probed through the loop-induced dipole in both the split and unsplit realisations (Sec.~\ref{subsec:ew_doublet_dipole}). The $f_{\rm required}$ maps size the gaps in that programme, but they do not prescribe how to close them. $f_{\rm required}$ is defined at fixed spectral shape, and $\Lambda_{90}$ grows only as $f^{1/8}$ for the dipole and scalar operators and $f^{1/12}$ for the dimension-7 family, so raw exposure is the weakest of the three levers available. The other two follow from the band-dependence argument of Sec.~\ref{subsec:halo_results}: moving $\omega_{\max}$ upward closes the wedge--contour gap for every Rayleigh operator, making TeV facilities such as CTA and SWGO~\cite{CTAConsortium2019Science,SWGOScience2025} the favourable direction there, while for the dipoles the gap is band-invariant and improved fractional precision is the only lever that acts.

%==================================================================
\section{Discussion and Conclusions}
\label{sec:conclusions}

We have revisited the photon--DM scattering calculation of
Ref.~\cite{acar2026darkmatterredblue} with three goals.

First, the two UV-complete channels considered there remain unobservable in any
astrophysical environment once perturbative validity is imposed. The minimal
dark-Higgs extension decouples the dark Yukawa from the SM vacuum expectation
value, but the perturbativity bound on $y_\chi$ and the LHC mixing-angle bound on
$\sin\theta$ together cap the effective coupling many orders of magnitude below
the \textit{Fermi}--LAT detection threshold. The gravitational channel is limited
by the smallness of $G_N$.

Second, we lifted the analysis out of those completions through a
model-independent operator scan, which begins from a known property of the basis
with an unremarked consequence for this channel. The $\partial^{\nu}F_{\mu\nu}$
vertex of the dimension-6 anapole and charge-radius operators vanishes for
on-shell photons~\cite{Zeldovich1957}, so attenuation is blind to them at tree
level at any exposure and whatever the dataset, and the real-photon basis reduces
to the dipoles and the Rayleigh family. Among the surviving basis, the dipoles enter
through two insertions and reach $\Lambda \simeq 0.32~\mathrm{GeV}$, flat at
$0.29~\mathrm{GeV}$ below $m_\chi \sim 1~\mathrm{GeV}$; the dimension-7 Rayleigh
operators reach $\Lambda \sim 0.79$--$1.06~\mathrm{GeV}$ and the dimension-6
scalar Rayleigh $\Lambda \simeq 0.21~\mathrm{GeV}$. 
All three families sit below
the EFT-validity wedge across the CDM mass range, by $0.30$--$0.64$~dex near
the CDM floor, and enter it only below $240~\mathrm{keV}$ for the dipoles,
$12$--$19~\mathrm{keV}$ for the dimension-7 Rayleigh operators and
$130~\mathrm{keV}$ for the scalar Rayleigh, where thermalisation with the photon
bath closes the window in any standard thermal history.

Third, we derived the first operator-resolved sensitivity estimates for
photon--DM scattering on astrophysical data, using a pixel-level reanalysis of 17~years of
\textit{Fermi}--LAT Pass~8 flux towards the Galactic
centre~\cite{StenhouseGhagDeppisch2026Totani}, with the sensitivity threshold
calibrated against pseudo-experiments rather than assumed. An independent
\textit{Fermi}--LAT dataset~\cite{Ackermann2015IGRB} reproduces the halo bound
within a factor of $1.4$ through a different removal channel, making
the agreement a test rather than a repetition. A PPPC~\cite{Cirelli2011PPPC}
annihilation template places its contour within a factor of a few of the
measured-template result, but on a fit too poor to claim the source-model
dependence as tested.

A statement sharper than EFT-invalidity follows for the dipoles
(Sec.~\ref{subsec:halo_results}): the charged mediator that would generate the
excluded moments radiatively must be lighter than
$\mathcal{O}(10)~\mathrm{MeV}$, far below the LEP floor on any new charged state,
so the excluded region contains no perturbative UV completion. Photon
attenuation in the Galactic halo does not yet constrain realisable parameter
space for any operator in the basis. What the analysis delivers is the
machinery, the operator-resolved structure of the channel, and a quantitative
statement of how far away observability is. The framework is written to be
reused, and extends to any instrument. Given a per-bin spectrum with
uncertainties and a line-of-sight column density, the pipeline recomputes sensitivity contours
without modification, so a new gamma-ray measurement can be used to derive
operator-resolved sensitivity estimates without rebuilding the machinery, and a
model builder can read off which operators the channel can see before
committing to a completion. Only the exposure multiplier $f_{\rm required}$ is
tied to the band it was computed in, since a change of energy band changes both
the spectral shape and the validity wedge.

The three UV translations of Sec.~\ref{sec:uv_translations} map that structure
onto the halo-preferred model space. The dark-Higgs portal probes the mediator
sector and requires a non-perturbative coupling to be seen. The kinetically mixed
dark photon is structurally invisible, its matched operators being those identified as vanishing, so the $(\epsilon, M_{A\p})$ plane is left to accelerator probes. The
electroweak-doublet dipole probes the annihilator, and is the case where this
channel earns its place: direct detection sits within a factor of $2.9$ of the
predicted moment, so the unsplit candidate is close to exclusion, and the
inelastic splitting that keeps it viable is precisely what closes direct
detection kinematically and the CMB annihilation bound thermally. Attenuation is
indifferent to that splitting and is the only one of the three that still
applies there.

Three questions remain open. The two-component sector, in which a heavy
annihilator sources the halo while a light scatterer redistributes its photons,
is naturally realised by a Rayleigh-coupled mediator such as the dark-Higgs
portal, since a pure kinetically mixed dark photon switches this channel off; the
halo-side quantification is the subject of
Ref.~\cite{StenhouseGhagDeppisch2026Totani}. Recent UV realisations of the
20~GeV excess~\cite{Yamashita2026TotaniHiggsPortal,Jho2025TotaniDilaton,NomuraTotani2026Higgsino,Yoshimatsu2026SecludedWW,Murayama2025BreitWigner}
span mediator masses from the sub-GeV range upward but confront the AMS-02 antiproton
exclusion~\cite{WangDuan2026AntiprotonExclusion} in different ways, and because
that bound is set in the same halo at the same velocity as the gamma-ray excess,
the velocity-dependent enhancements those models invoke do not by themselves
relieve it; a joint gamma-ray and antiproton analysis of the window is the
natural next test. Finally, the unpolarised optical depth used throughout
discards the linear polarisation of the scattered photons, which distinguishes
the spin-2 graviton exchange from the scalar and dipole operators and is a
handle no unpolarised measurement can reproduce.

%------------------------------------------------------------------

\section*{Acknowledgments}
T.R.S. acknowledges support from an STFC studentship (grant number 187357). C.G. acknowledges support from the UK Science and Technology Facilities Council (STFC) via the Consolidated Grant UKRI2851. F. F. D. acknowledges support from STFC via the Consolidated Grant ST/X000613. A.A., M.B. and D.P.W. acknowledge support from STFC via the Consolidated Grant ST/Y000285/1. We acknowledge Cameron J. Farquhar for helpful suggestions in editing. This work used \textit{Fermi}--LAT data provided by the Fermi Science Support Center, the \texttt{emcee} package~\cite{emcee2013}, and PPPC~4~DM~ID spectra~\cite{Cirelli2011PPPC}. 

\section*{Data Availability}
The analysis code and figure-generation scripts that
reproduce every figure in this paper are archived on Zenodo at
\url{https://doi.org/10.5281/zenodo.21703809}.
The underlying \textit{Fermi}--LAT Pass~8 photon data are public and available from the
Fermi Science Support Center. Information regarding data availability for the pixel-level Galactic-centre halo posterior that
drives the halo bounds is released with
Ref.~\cite{StenhouseGhagDeppisch2026Totani}. The remaining external inputs are public at
their original sources: the \textit{Fermi}--LAT isotropic diffuse gamma-ray
background~\cite{Ackermann2015IGRB}, the PPPC~4~DM~ID annihilation
templates~\cite{Cirelli2011PPPC}, and the dwarf spheroidal data%
~\cite{McDaniel2024DsphLegacy,GeringerSameth2015Dfactors}.

\bibliography{bib}
\end{document}